\documentclass{aifrontiers}
\usepackage{aifrontiers}
\usepackage{booktabs}
\usepackage{array}
\usepackage{colortbl}
\usepackage{caption}
\usepackage{subcaption}
\usepackage{multirow}
\usepackage{multicol}
\usepackage{float}
\usepackage{enumitem}
\usepackage{xspace}
\usepackage{soul}
\usepackage{listings}
\usepackage[export]{adjustbox}
\usepackage{algorithm}
\usepackage{algorithmic}

\usepackage[T1]{fontenc}    %
\usepackage[utf8]{inputenc} %

\newlength\savewidth

\newcolumntype{x}[1]{>{\centering\arraybackslash}p{#1pt}}
\newcolumntype{y}[1]{>{\raggedright\arraybackslash}p{#1pt}}

\definecolor{baselinecolor}{gray}{0.93}

\definecolor{bluebg}{RGB}{225, 235, 255}
\definecolor{bluetext}{RGB}{0, 0, 139}
\definecolor{orangebg}{RGB}{255, 228, 196}
\definecolor{orangetext}{RGB}{255, 69, 0}
\definecolor{purplebg}{RGB}{230, 230, 250}
\definecolor{purpletext}{RGB}{128, 0, 128}
\definecolor{redbg}{RGB}{255, 204, 204}
\definecolor{redtext}{RGB}{178, 34, 34}
\definecolor{greenbg}{RGB}{230, 255, 230}
\definecolor{greentext}{RGB}{0, 100, 0}
\definecolor{cyanbg}{RGB}{224, 255, 255}
\definecolor{cyantext}{RGB}{0, 139, 139}
\definecolor{yellowbg}{RGB}{255, 255, 204}
\definecolor{yellowtext}{RGB}{205, 133, 63}
\definecolor{violetbg}{RGB}{245, 230, 255}
\definecolor{violettext}{RGB}{148, 0, 211}
\definecolor{teal}{HTML}{1E88E5}
\definecolor{coral}{HTML}{D81B60}

\newcommand{\cellcolorpercent}[2]{
    \ifdim #1pt < #2pt
        \cellcolor{coral!15}#1\%
    \else
        \cellcolor{teal!15}#1\%
    \fi
    & 
    \ifdim #2pt < #1pt
        \cellcolor{coral!15}#2\%
    \else
        \cellcolor{teal!15}#2\%
    \fi
}

\newtcolorbox{quotefigurebox}{
  colframe=gray, colback=white, arc=5mm,
  boxrule=0.5pt, left=1pt, right=1pt, top=1pt, bottom=1pt,
  sharp corners,
}

\definecolor{rescolor}{RGB}{180,40,40}

\newcommand{\agentlens}{\textsc{AgentLens}\xspace}

\begin{document}

\title{\agentlens: Revealing The Lucky Pass Problem in SWE-Agent Evaluation}

\shorttitle{\agentlens: Revealing The Lucky Pass Problem in SWE-Agent Evaluation}

\author{
    {\bfseries Priyam Sahoo$^{1,2,}$\thanks{Work done as a student researcher while at Microsoft.}, Gaurav Mittal$^{2,}$\thanks{Co-second authors.}, Xiaomin Li$^{2,}$\footnotemark[2], Shengjie Ma$^{2}$, Benjamin Steenhoek$^{2}$,\\ Pingping Lin$^{2}$, Yu Hu$^{2}$} \\[0.45em]
    {\normalsize $^{1}$University of Illinois, Urbana-Champaign, IL, USA \quad $^{2}$Microsoft, Redmond, WA, USA}\\[0.45em]
    {\normalsize \texttt{priyams3@illinois.edu}, \quad \texttt{Gaurav.Mittal@microsoft.com}}
}
\date{\today}
\renewcommand{\thefootnote}{\fnsymbol{footnote}}

\renewcommand{\weblink}{}
\renewcommand{\foundrylink}{}
\renewcommand{\hflink}{https://github.com/agentlens-framework/agentlens}
\renewcommand{\ghlink}{}

\begin{abstract}
Evaluation of software engineering (SWE) agents is dominated by a binary signal: whether the final patch passes the tests. This outcome-only view treats a principled solution and a chaotic trial-and-error process as equivalent. We show that this equivalence is empirically false. We evaluate 2{,}614 OpenHands trajectories from eight model backends on 60 SWE-bench Verified tasks. Of these, 47 have enough passing trajectories to construct task-level process references, yielding a 1{,}815-trajectory evaluation subset. Among passing trajectories in this subset, 10.7\% exhibit behavior we call a \emph{Lucky Pass}: regression cycles, blind retries, missing verification, or temporally disordered exploration, implementation, and verification.
We introduce \textbf{\agentlens}, a framework for process-level assessment of SWE-agent trajectories, and release \textbf{\agentlens-Bench}, a dataset of 1{,}815 trajectories annotated with quality scores, waste signals, divergence points, and 47 task-level Prefix Tree Acceptor (PTA) references. \agentlens builds PTA references by merging multiple passing solutions for the same task, and uses a context-sensitive intent labeler to assign actions to \textit{Exploration}, \textit{Implementation}, \textit{Verification}, or \textit{Orchestration} based on trajectory history rather than tool identity alone.
On \agentlens-Bench, the quality score separates passing trajectories into Lucky, Solid, and Ideal tiers and further decomposes Lucky Passes into five recurring mechanisms. Across the eight model backends, Lucky rates range from 0.5\% to 23.2\%, and some models move by as many as five rank positions when ranked by quality score instead of pass rate.
We publicly release \agentlens-Bench and the \agentlens SDK at \url{https://github.com/agentlens-framework/agentlens}, with a project page at \url{https://agentlens-framework.github.io/agentlens/}.
\end{abstract}

\maketitle

% ============================================================
%  1  INTRODUCTION
% ============================================================
\section{Introduction}
\label{sec:intro}

Software engineering agents have moved quickly from prototypes to systems that resolve real GitHub issues end-to-end. SWE-agent~\citep{yang2024swe}, OpenHands~\citep{wang2025openhands}, AutoCodeRover~\citep{zhang2024autocoderover}, Agentless~\citep{xia2024agentless}, and Devin~\citep{devin} all read codebases, edit files, and run test suites without human input. The benchmark anchoring this progress, SWE-bench~\citep{jimenez2024swe}, evaluates these systems with a binary signal: does the final patch pass the tests? That signal is useful for measuring capability, but insufficient for evaluating behavior. Consider two agents resolving the same issue. One explores the repository in a few targeted steps, identifies the root cause, applies a minimal fix, and verifies it. The other repeatedly attempts similar edits, loops through failed checks, and eventually reaches a working patch through trial and error. Both receive the same SWE-bench label of ``resolved.'' The behavioral difference is real, important for downstream uses of trajectories, and invisible to outcome-only evaluation.

We show that this conflation occurs in practice. Across 1{,}136 passing agent trajectories from eight model backends on SWE-bench Verified, 10.7\% are reached through behavior we call a \emph{Lucky Pass}: regression cycles, blind retries, missing verification, or temporally disordered exploration, implementation, and verification. A further 69.1\% are Solid but imperfect, and only 20.2\% are Ideal: principled, low-waste, and well-ordered. Pass-rate rankings disagree with process-quality rankings on all eight configurations, and Lucky rates range from 0.5\% to 23.2\% across models.

\begin{figure}[t]
  \centering
  \includegraphics[width=0.8\linewidth]{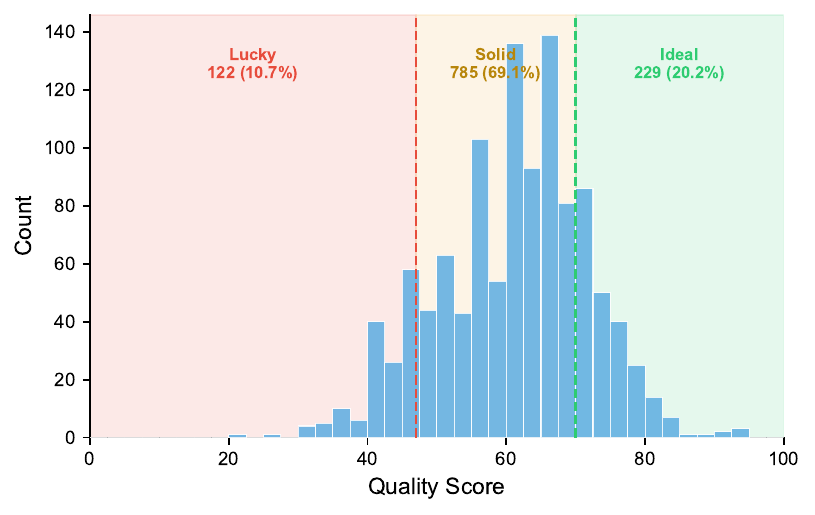}
%   \vspace{-1.5em}
  \caption{\textbf{Passing trajectories are not behaviorally homogeneous.}
  Among 1{,}136 passing trajectories in \agentlens-Bench, \agentlens classifies 229 as Ideal (20.2\%), 785 as Solid (69.1\%), and 122 as Lucky (10.7\%). Binary evaluation treats all of these trajectories as equally successful, while process-aware scoring separates direct, coherent solutions from weak processes that happen to pass.}
  \label{fig:quality-distribution}
%   \vspace{-2em}
\end{figure}

This matters for three reasons. First, trajectory datasets such as SWE-Gym~\citep{pan2025swe}, R2E-Gym~\citep{jain2025r2e}, and SWE-smith~\citep{yang2025swesmith} commonly filter on pass rate, treating every successful trajectory as equally valuable supervision. This makes pass-rate filtering a coarse proxy for demonstration quality: a trajectory that reaches the correct outcome through brittle exploration or excessive retry is selected in the same way as a direct, coherent solution. Second, as models converge on pass-rate benchmarks, process quality becomes a useful axis for model comparison. In Section~\ref{sec:lucky-taxonomy} and Table~\ref{tab:model-comparison}, we show that ranking models by \agentlens quality score changes their ordering relative to pass rate, with some models moving by as many as five rank positions. Third, deployment risk depends on process. Agents that succeed by repeated trial and error may behave unpredictably when repositories are large, tests are expensive, or actions are irreversible.

\begin{figure}[t]
  \centering
  \includegraphics[width=\linewidth]{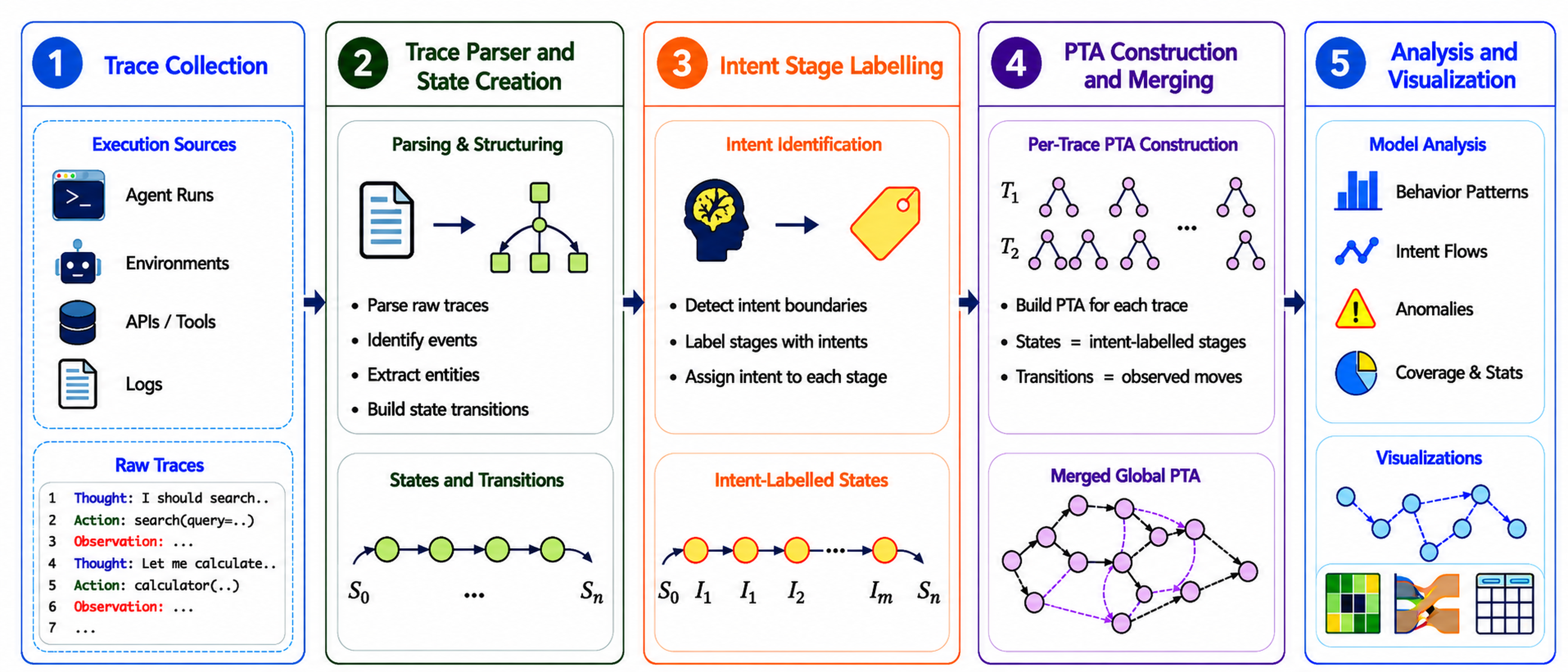}
  \caption{\textbf{\agentlens workflow.} \agentlens starts from raw execution traces and converts them into structured state sequences. Intent-stage labeling assigns each state one of four process roles: Exploration, Implementation, Verification, or Orchestration. These intent-labeled states become the states used for per-trace PTA construction. Passing trajectories for the same task are then merged into a task-level PTA, which represents the known-good solution strategies for that task. New trajectories are scored against this task-level reference to produce quality tiers, divergence points, waste reports, and visual analyses.}
  \label{fig:agentlens}
%   \vspace{-2em}
\end{figure}

We introduce \agentlens, a framework for process-level assessment of SWE-agent trajectories (Figure~\ref{fig:agentlens}). \agentlens has two technical components. The first is a PTA-based quality reference: instead of comparing a candidate trajectory to a single reference trace, we build a Prefix Tree Acceptor~\citep{oncina1992inferring} from multiple passing solutions for the same task.
% The resulting directed acyclic graph encodes a space of known-good strategies, allowing valid alternative paths to be distinguished from off-path behavior.
The resulting directed acyclic graph encodes a space of known-good strategies. This lets \agentlens recognize valid alternative solution paths while flagging redundant retries, irrelevant exploration, and divergence from known successful processes.
The second is context-sensitive intent-stage labeling. Each action is assigned to one of four cognitive phases: \textit{Exploration}, \textit{Implementation}, \textit{Verification}, or \textit{Orchestration}, using trajectory history rather than tool identity alone. This resolves common ambiguities such as terminal commands: \texttt{grep} remains exploratory even after an edit, whereas \texttt{pytest} is verification.

Together, these steps give \agentlens a task-specific reference for judging not only whether an agent solved a task, but how it moved through the solution process. Each raw trajectory is first converted into an intent-labeled state sequence. Passing trajectories for the same task are then merged into a task-level PTA, and new trajectories are scored against that PTA to compute a composite quality score, tier label, divergence point, and structured inefficiency report. We validate the intent-stage labels with a seven-annotator agreement study, obtaining Fleiss' $\kappa=0.933$, and evaluate the scoring pipeline on 2{,}614 trajectories from 60 SWE-bench Verified tasks. Of these 60 tasks, 47 have enough passing trajectories to build a task-level PTA. These 47 tasks form \textbf{\agentlens-Bench}, which contains 1{,}815 process-annotated trajectories with quality scores, waste annotations, divergence metadata, and one ground-truth PTA per task. Because each PTA is tied to a distinct SWE-bench task, this collection provides task-diverse references for scoring trajectories, filtering training pools, and analyzing failures across different solution spaces.

Below are our main contributions:
\begin{itemize}[leftmargin=1.5em, topsep=4pt, itemsep=3pt]
  \item \textbf{\agentlens-Bench.} To our knowledge, \agentlens-Bench is the first process-annotated SWE-agent trajectory dataset. It contains 1{,}815 trajectories from 47 PTA-eligible SWE-bench Verified tasks, with 40-column feature vectors, one task-level ground-truth PTA per task, waste annotations, divergence metadata, and tier labels. Because each PTA represents the known-good solution space for a distinct task, the collection provides task-diverse references for trajectory scoring, filtering, failure analysis, and future process-aware training studies.

  \item \textbf{The Lucky Pass finding and taxonomy.} We show that 10.7\% of passing trajectories reach correct patches through weak processes. These Lucky Passes decompose into five behavioral categories with significant model associations ($\chi^2(28)=102.47$, $p<0.0001$). Across the eight evaluated model backends, the share of successful trajectories classified as Lucky ranges from 0.5\% to 23.2\%.

  \item \textbf{Context-sensitive intent labeling.} We introduce a trajectory-history-aware labeler that resolves exploration-vs-verification ambiguity in terminal commands, validated at $\kappa=0.933$ on 200 states with 96.0\% raw agreement across seven annotators.

  \item \textbf{PTA-based process references and quality scoring.} We introduce a PTA-based representation that merges passing trajectories for the same task into a task-level reference of known-good solution strategies. \agentlens then scores new trajectories by combining structural alignment with coverage, coherence, and temporal-profile signals. On \agentlens-Bench, this combined score significantly separates passing from failing trajectories ($p=0.0017$).

  \item \textbf{Open-source tooling.} We publicly release an SDK and web interface for process-aware trajectory analysis at \url{https://github.com/agentlens-framework/agentlens}. The tooling supports ATIF trajectory logs~\citep{harbor_atif} and OpenHands traces~\citep{wang2025openhands}. The released scoring pipeline is deterministic and does not require LLM calls or external API access. Because \agentlens operates on intent-labeled states rather than raw logs, trajectories from other scaffolds can be analyzed after conversion to ATIF or through a lightweight trace adapter. Our experiments validate the full pipeline on OpenHands trajectories; other scaffolds may need updates to the tool registry and command-level rules before their scores are comparable.
\end{itemize}

% ============================================================
%  2  RELATED WORK
% ============================================================
\section{Related Work}
\label{sec:related}

\paragraph{Outcome-based SWE-agent benchmarks.}
SWE-bench~\citep{jimenez2024swe} established binary pass/fail as the standard for coding-agent evaluation, and subsequent benchmarks refine this outcome signal through human validation, decontamination, live issue streams, multi-language coverage, or realistic task pricing~\citep{openai2024swebenchverified,badertdinov2025swe,zhang2025swebenchlive,wang2025multiswe,miserendino2025swelancer}. Adjacent code benchmarks such as LiveCodeBench~\citep{jain2024livecodebench}, BigCodeBench~\citep{zhuo2024bigcodebench}, and TerminalBench~\citep{stanford2025terminalbench} similarly evaluate final correctness. ABC~\citep{zhu2025abc} documents measurement errors in this benchmark family, including insufficient test coverage. These works improve outcome evaluation; \agentlens instead measures the process that produced the outcome.

\paragraph{Process-level trajectory evaluation.}
Graphectory~\citep{liu2026process} is the closest prior work: it encodes execution traces as graphs and computes process-centric metrics independently of task success. Closest to our PTA-based reference, recent work merges passing traces and validates new executions against them for UI agents~\citep{PTA_sharma2026}. Other studies characterize successful and failing SWE-agent trajectories through thought-action-result patterns, length, variance, or patch quality~\citep{bouzenia2025understanding,majgaonkar2025understanding}, while TRAIL~\citep{deshpande2025trail}, Agent-as-a-Judge~\citep{zhuge2024agentasjudge}, AgentBoard~\citep{ma2024agentboard}, and AgentBench~\citep{liu2024agentbench} study broader agent evaluation beyond final success. Process reward models and step-level supervision make a related argument that intermediate reasoning signals differ from outcome labels~\citep{uesato2022solving,lightman2023verify,wang2024mathshepherd,zheng2025processbench,pan2025webshepherd,shum2025swerm}. \agentlens differs by providing deterministic, decomposable process scores for SWE trajectories, using context-sensitive intent labels, PTA references built from multiple passing solutions, and structured inefficiency attribution with divergence localization.

\paragraph{Trajectory datasets for SWE agents.}
SWE-Gym~\citep{pan2025swe}, R2E-Gym~\citep{jain2025r2e}, SWE-smith~\citep{yang2025swesmith}, and OpenHands logs~\citep{wang2025openhands} provide execution traces or training instances, but they filter or organize trajectories primarily by outcome. To our knowledge, no released coding-agent dataset provides per-trajectory quality scores, ground-truth reference graphs, divergence localization, and waste annotations together. \agentlens-Bench fills this gap. Appendix~\ref{app:extended-related} provides compact dataset and framework comparison tables.

% ============================================================
%  3  HOW AGENTLENS WORKS
% ============================================================
\section{How \agentlens Works}
\label{sec:method}

\agentlens evaluates a candidate trajectory in four stages: it parses raw logs into labeled states, constructs a task-specific reference graph from passing solutions, scores the candidate against that reference, and returns a structured quality report.

\subsection{From raw logs to labeled states}
\label{sec:parsing}

An agent log is a sequence of tool calls paired with environment responses. \agentlens parses each step into a state containing the tool, target file, affected line range, content hash, trajectory position, and an intent-stage label. We use four cognitive phases: \textbf{Exploration} (E; reading files, searching, listing directories), \textbf{Implementation} (I; editing or creating source files), \textbf{Verification} (V; running tests, checking errors, re-reading edited files), and \textbf{Orchestration} (O; bookkeeping and reasoning steps). These phases follow empirical studies of developer cognition~\citep{ko2006exploratory,alaboudi2021exploratory} and prior trajectory analysis~\citep{liu2026process}.

A key challenge is that tool identity alone is insufficient. For example, \texttt{read\_file(test\_api.py)} may be exploratory before any patch is written but verifying after an implementation step. We therefore use a deterministic, rule-based, context-sensitive labeler that tracks whether implementation has occurred and which files have been edited. Search and file-inspection commands such as \texttt{grep}, \texttt{cat}, and \texttt{ls} are labeled E, test-running commands such as \texttt{pytest} are labeled V, source edits are labeled I, and reads of previously edited files are labeled V. The full registry and rule decision flow are provided in Appendices~\ref{app:tool-registry} and~\ref{app:intent-flow}. Section~\ref{sec:rq-labels} reports the reliability study that validates these labels before they are used for scoring.

\subsection{Building a PTA reference}
\label{sec:pta}

\begin{figure}[!ht]
  \centering
  \includegraphics[width=\linewidth]{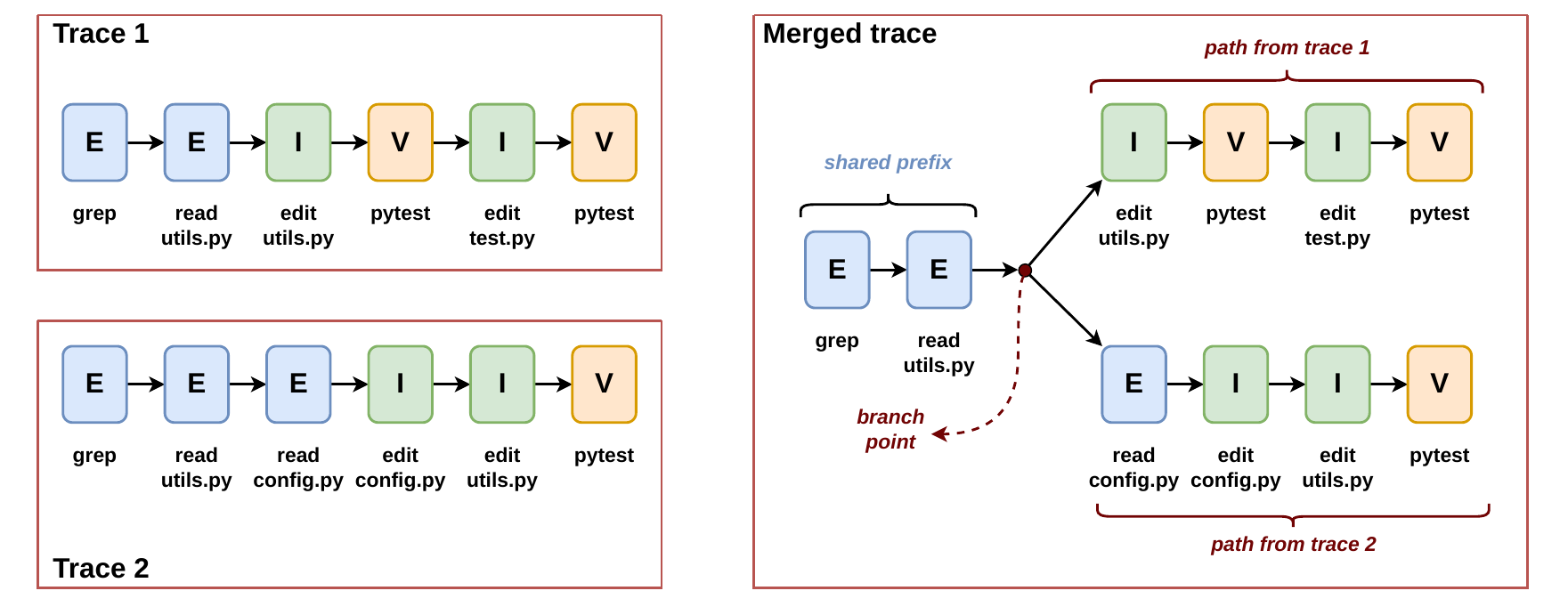}
%   \vspace{-2em}
  \caption{\textbf{PTA construction.} Two individual passing traces sharing an early exploration prefix but diverging at implementation are merged into a single DAG. Shared nodes reflect equivalent actions across agents; branches reflect genuine strategic divergence.}
  \label{fig:pta-construction}
%   \vspace{-1em}
\end{figure}

A single reference trajectory cannot represent the diversity of correct strategies: two agents may solve the same task through different but valid sequences. \agentlens instead constructs a Prefix Tree Acceptor (PTA)~\citep{oncina1992inferring} from $k\geq 2$ passing trajectories for the same task. Any passing trajectory of the task can be merged into its PTA, and the trajectory being scored is never part of its own PTA (Section~\ref{sec:setup}); Appendices~\ref{app:robust-lomo} and~\ref{app:robust-reference-quality} examine how the choice of merged trajectories affects the scores.
This approach of merging multiple passing executions into a PTA and validating new runs against it, while tolerating valid variation, follows recent work on learning correct agent behavior from passing traces across domains such as UI agents~\citep{PTA_sharma2026}. We adapt it to SWE-agent trajectories with intent-stage labeling and a process-quality score.
Shared prefixes are merged into common nodes, while divergent but successful strategies form branches. Each root-to-terminal path therefore represents one known-good solution, and the resulting directed acyclic graph encodes a space of correct behaviors rather than a single exemplar (Figure~\ref{fig:pta-construction}).

During construction, states from different trajectories are merged when they represent equivalent actions. The equivalence engine handles surface variation such as different tool names, overlapping file regions, and equivalent terminal commands. For example, \texttt{grep} and \texttt{rg} calls with the same search intent can match the same PTA state rather than being treated as different actions. Appendix~\ref{app:equivalence} gives the full equivalence cascade and thresholds.

\subsection{Scoring a candidate trajectory}
\label{sec:scoring}

Given a candidate trajectory $\tau_c$ and task PTA $\mathcal{G}$, \agentlens computes four complementary signals. \textbf{Structural alignment} ($\Phi_{\mathrm{struct}}$) measures whether the candidate visits PTA states in roughly the right order, combining ordered recall with unordered precision. \textbf{Set coverage} ($\Phi_{\mathrm{cov}}$) measures the fraction of PTA states matched by the candidate regardless of order. \textbf{Trajectory coherence} ($\Phi_{\mathrm{coh}}$) summarizes the intent-stage sequence, rewarding forward progress such as E$\rightarrow$I$\rightarrow$V and penalizing backtracks and blind retries. \textbf{Temporal profile similarity} ($\Phi_{\mathrm{temp}}$) compares the candidate's stage distribution over early, middle, and late trajectory segments against the PTA using Jensen-Shannon divergence~\citep{lin1991divergence}. Intuitively, the first two signals ask whether the agent touched the right parts of the solution space, while the latter two ask whether it moved through them in a plausible problem-solving order. Appendix~\ref{app:scoring} provides formal definitions and a worked scoring example.

The four signals are combined into a 0--100 quality score:
\begin{equation}
  f(\tau_c,\mathcal{G}) =
  0.20\cdot\Phi_{\mathrm{struct}}
  + 0.15\cdot\Phi_{\mathrm{cov}}
  + 0.30\cdot(100\cdot\Phi_{\mathrm{coh}})
  + 0.35\cdot(100\cdot\Phi_{\mathrm{temp}}).
  \label{eq:combined}
\end{equation}
$\Phi_{\mathrm{struct}}$ and $\Phi_{\mathrm{cov}}$ are percentage-based, while $\Phi_{\mathrm{coh}}$ and $\Phi_{\mathrm{temp}}$ are rescaled from $[0,1]$. The weights were selected by grid search to maximize pass/fail AUROC on a disjoint pilot calibration set (Section~\ref{sec:setup}) and then held fixed for all scaled experiments. Behavioral signals receive 65\% of the total weight, reflecting that structural coverage and process quality fail on different trajectories.

\subsection{Reports, tiers, and waste signals}
\label{sec:report}

The final report includes the composite score, per-stage coverage, divergence-point localization, and structured inefficiency analysis. Waste is detected in five categories: regression loops, blind retries, redundant steps, unnecessary exploration, and cyclic patterns. Each instance is localized to trajectory steps and attributed to tools where possible. For downstream analysis, we use fixed quality tiers. Tiers are assigned on the report's quality score $q = 0.25\cdot\mathrm{Cov} + 0.25\cdot(100\cdot\Phi_{\mathrm{coh}}) + 0.18\cdot(100\cdot\mathrm{SC}) + 0.12\cdot(100\cdot\mathrm{WS}) + 0.10\cdot\Phi_{\mathrm{struct}} + 0.10\cdot O$, where Cov is coverage of the best-matching PTA path, SC (stage completeness) is the fraction of the reference's intent stages matched at least once, WS (workflow similarity) is the longest-common-subsequence ratio between the candidate's and the reference's intent-stage sequences, and $O$ is 100 for passing and 0 for failing trajectories. The outcome term shifts all trajectories with the same outcome equally, so it does not change their relative order. The combined score $f$ above is used for pass/fail discrimination (Section~\ref{sec:complementarity}). Passing trajectories with score $\geq 70$ are \textbf{Ideal}, those with $47\leq\mathrm{score}<70$ are \textbf{Solid}, and those with score $<47$ are \textbf{Lucky}. Failing trajectories are labeled \textbf{Partial-fail} when score $\geq 40$ and \textbf{Off-track} otherwise. The Lucky cutoff is the pass/fail decision threshold of the combined score $f$ selected on the pilot calibration set (Youden's J, 46.4; Appendix~\ref{app:score-distributions}), rounded up to 47 and applied to $q$. Because the outcome term adds 10 points to every passing trajectory, a passing trajectory is Lucky when its score without that term is below 37. The Ideal cutoff of 70 is a reporting convention that separates the strongest trajectories from the Solid group rather than a second optimized boundary. Both thresholds were held fixed for the scaled evaluation.
Additional report fields and the five-level verdict are described in Appendix~\ref{app:scoring}.
% Additional report fields, the five-level verdict, and the planned SDK/web interface are described in Appendix~\ref{app:web-app}.

% ============================================================
%  4  EXPERIMENTAL SETUP
% ============================================================
\section{Experimental Setup}
\label{sec:setup}

\paragraph{Compute environment.}
All \agentlens scoring, PTA construction, stratification, and waste analysis experiments were run locally on a machine with 11 CPU cores and 18GB memory. The pipeline is CPU-only and does not require GPU workers; trajectory generation uses external model API calls and is separate from the post-hoc \agentlens analysis reported here.

\paragraph{Dataset.}
We evaluate \agentlens on trajectories generated by the OpenHands coding agent~\citep{wang2025openhands} on SWE-bench Verified~\citep{openai2024swebenchverified}. The corpus contains 2{,}614 trajectories across 60 tasks and eight model backends: GPT-4.1, GPT-4o, GPT-5.2-Codex, GPT-5.3-Codex, Claude Sonnet 4.5, Claude Opus 4.5, Claude Opus 4.6, and Gemini 2.5 Pro~\citep{deepmind2024gemini}. Across these trajectories, 1{,}389 pass, 1{,}217 fail, and 8 have unrecorded outcomes. PTA construction is task-specific and requires multiple passing trajectories for a task, since a merged PTA is built from multiple known-good solutions. We require at least five, the merge count used for the scaled evaluation (Section~\ref{sec:ablation}). This requirement is satisfied by 47 of the 60 tasks, spanning 1{,}815 trajectories: 1{,}136 passing and 679 failing. All scoring, stratification, and waste analysis is performed on this 1{,}815-trajectory subset, which constitutes the \agentlens-Bench release.

\paragraph{Calibration and holdout.}
Signal weights were calibrated on a separate pilot set of 278 trajectories across 10 tasks. The pilot was used only for grid-search weight optimization (step 0.05, unit-sum constraint, AUROC-maximizing), yielding $w=(0.20,0.15,0.30,0.35)$ with pilot AUROC $=0.755$ and pilot F1 $=0.791$. No pilot trajectories appear in the scaled evaluation set, and the weights are frozen for all subsequent experiments. In the scaled set, PTA construction is task-specific: for each scored trajectory, the reference PTA is built from other passing trajectories for the same task, excluding the trajectory being scored. This lets us assign quality scores to all 1{,}136 passing trajectories without scoring any trajectory against a PTA that contains itself. Pass/fail discrimination is computed entirely on the scaled set with no pilot data.

\paragraph{Baselines.}
We compare against three reference strategies: individual trajectory matching, which scores each test trajectory against every passing training trajectory and reports the best match; TF-IDF alignment in the BERTScore style~\citep{zhang2020bertscore}; and dense embedding alignment with \texttt{text-embedding-3-large}.

\paragraph{Metrics.}
We report micro-averaged AUROC~\citep{fawcett2006introduction} as the primary discrimination metric because task-level class balance varies across evaluation slices. Decision thresholds are selected by Youden's J~\citep{youden1950index}. For significance testing, we report the Kolmogorov--Smirnov test $p$-value comparing passing and failing score distributions.

% ============================================================
%  5  RESULTS
% ============================================================

\section{Results}
\label{sec:results}

We organize the results around the main empirical findings first, followed by validation checks. Section~\ref{sec:stratification} shows that passing trajectories are not behaviorally homogeneous: 10.7\% are Lucky Passes despite producing correct patches. Section~\ref{sec:lucky-taxonomy} analyzes these Lucky Passes, decomposes them into five weak-success mechanisms, and shows how process quality changes model comparison. Section~\ref{sec:complementarity} tests whether the resulting quality score also separates passing and failing trajectories. Section~\ref{sec:rq-labels} validates the intent-stage labels used to construct trajectory states.

\subsection{Passing trajectories are not behaviorally homogeneous}
\label{sec:stratification}

The central question for \agentlens is whether all successful trajectories should be treated as equally good demonstrations. Across 1{,}136 passing trajectories eligible for assessment, the answer is no. Applying the fixed tier thresholds from Section~\ref{sec:report} yields 229 Ideal trajectories (20.2\%), 785 Solid trajectories (69.1\%), and 122 Lucky trajectories (10.7\%). Figure~\ref{fig:quality-distribution} shows this distribution. Binary evaluation assigns all 1{,}136 trajectories the same label, while \agentlens separates direct, coherent solutions from weak processes that happen to pass.

Not all non-Ideal trajectories are weak in the same way. Some have low structural overlap with the merged PTA but remain coherent and temporally well organized. We call this profile \emph{efficient-but-atypical}: the agent follows an unconventional but valid path rather than the dominant known-good branches. This distinction matters because low structural overlap alone would make these trajectories look weak, while the full \agentlens score separates them from Lucky Passes. Appendix~\ref{app:behavioral-profiles} reports the full four-profile breakdown.

\subsection{Lucky Passes reveal recurring weak-success mechanisms}
\label{sec:lucky-taxonomy}

We now analyze the 122 Lucky Passes: trajectories that produce correct patches but receive low process-quality scores. Using automatically computed \agentlens signals, including trajectory length, verification coverage, waste patterns, implementation coverage, and coherence, we assign each Lucky Pass to one of five mutually exclusive categories (Table~\ref{tab:lucky-taxonomy-main}). The two largest categories are C2, Brute-Force Convergence, where the agent reaches a correct patch through repeated low-coherence attempts, and C3, Incomplete Implementation, where the agent makes a partial fix that passes because visible tests are insufficient. Together, C2 and C3 account for 68.0\% of Lucky Passes.

\begin{table}[!ht]
\centering
\caption{\textbf{Lucky Pass taxonomy.} Five categories with discriminating signals, prevalence, and causal mechanisms. C2 and C3 together account for 68.0\% of all Lucky Passes.}
\label{tab:lucky-taxonomy-main}
\small
\begin{tabular}{@{}lcccl@{}}
\toprule
\textbf{Category} & \textbf{$n$} & \textbf{\%} & \textbf{Key Signal} & \textbf{Primary Cause} \\
\midrule
C1: Minimal \& Unverified     & 19 & 15.6 & Short + no V stage   & Agent overconfidence \\
C2: Brute-Force Convergence   & 42 & 34.4 & High waste, low coh.  & Lack of planning \\
C3: Incomplete Implementation & 41 & 33.6 & Partial fix, low cov. & Test-suite gaps \\
C4: Excessive Exploration     & 5  & 4.1  & Very long, unfocused  & Missing termination \\
C5: Divergent-but-Valid       & 15 & 12.3 & Alternative approach  & Multiple valid solutions \\
\bottomrule
\end{tabular}
\end{table}

The waste signals explain why these trajectories are weak despite passing. We track regression loops, blind retries, redundant steps, unnecessary exploration, and cyclic patterns as defined in Section~\ref{sec:report}. These signals are computed relative to the task-level PTA, so behavior already present in a known-good solution path is not counted as waste. Across pass/fail groups, unnecessary exploration and cyclic patterns are the strongest discriminators: failing trajectories are 58\% more likely to inspect files outside the known-good solution space (F/P $=1.58$), and cyclic patterns are 32\% more prevalent in failing trajectories (F/P $=1.32$). The full pass/fail waste breakdown is in Appendix~\ref{app:waste-breakdown}.

For Lucky Passes, the clearest waste pattern is not how often blind retries appear, but how much work they waste when they do appear. Lucky trajectories with blind retries waste 11.4 steps per instance, compared with 2.7 in Ideal trajectories, a 4.2$\times$ increase. This captures a common weak-success pattern: instead of systematically debugging a failure, the agent repeats similar actions until one attempt succeeds. Appendices~\ref{app:waste-tiers} and~\ref{app:failure-gallery} provide the full tier-wise waste table and representative timelines.

The Lucky categories are not evenly distributed across models. A chi-square test over model and category assignments shows a significant association ($\chi^2(28)=102.47$, $p<0.0001$). Opus 4.6 disproportionately produces Minimal-and-Unverified passes, GPT-4.1 produces most Brute-Force and Excessive-Exploration cases, and the Codex variants concentrate in Incomplete Implementation. Lucky Passes are also task-concentrated: the top 10 tasks account for 63.1\% of all Lucky Passes. Appendix~\ref{app:lucky-analysis} gives the decision tree, model and task cross-tabs, category-level waste analysis, verification-gap analysis, and extended case studies.

\paragraph{Model comparison.}
Process quality changes how models are ranked on \agentlens-Bench. Table~\ref{tab:model-comparison} reports model-level aggregates over the 1{,}815-trajectory PTA-eligible subset, with quality scores (QS) computed for each model's successful trajectories and compared against pass-rate (PR) rankings on the same subset. The rankings disagree for every model. GPT-4o rises from 8th by pass rate to 3rd by quality score, but this should be interpreted conditionally: among the tasks it solves, its successful trajectories tend to receive higher process-quality scores. Since GPT-4o succeeds less often, this shift may partly reflect which tasks it solves rather than a model-wide advantage. On the 22 tasks solved by all eight backends the ordering is unchanged (Appendix~\ref{app:robust-shared}), and its process quality remains third when its failing trajectories are included (Appendix~\ref{app:robust-per-attempt}). The quality score therefore complements pass rate rather than replacing it as a measure of overall capability. Opus 4.6 shows the opposite pattern: a 77.3\% pass rate hides an 18.7\% Lucky rate. Across models, the share of successful trajectories classified as Lucky ranges from 0.5\% to 23.2\%.

\begin{table}[!ht]
\centering
\caption{\textbf{Frontier model comparison.} Quality-score rankings (QS Rank) disagree with pass-rate rankings (PR Rank) on all eight models. Cond.\ Quality is the conditional quality score over each model's passing trajectories; Appendix~\ref{app:robust-per-attempt} reports quality over all attempts.}
\label{tab:model-comparison}
\small
\begin{tabular}{@{}lccccc@{}}
\toprule
\textbf{Model} & \textbf{Pass\%} & \textbf{PR Rank} & \textbf{Cond.\ Quality} & \textbf{QS Rank} & \textbf{Lucky\%} \\
\midrule
sonnet-4.5     & 86.8\% & 2 & \textbf{67.4} & 1 & 1.0\% \\
opus-4.5       & 87.9\% & 1 & 66.2          & 2 & 0.5\% \\
gpt-4o         & 34.9\% & 8 & 63.4          & 3 & 4.1\% \\
gemini-2.5-pro & 42.9\% & 7 & 59.2          & 4 & 7.6\% \\
gpt-5.3-codex  & 45.9\% & 6 & 58.3          & 5 & 15.3\% \\
opus-4.6       & 77.3\% & 3 & 56.7          & 6 & 18.7\% \\
gpt-5.2-codex  & 64.6\% & 4 & 56.1          & 7 & 19.4\% \\
gpt-4.1        & 59.9\% & 5 & 54.7          & 8 & \textbf{23.2\%} \\
\bottomrule
\end{tabular}
\vspace{-10pt}
\end{table}

Additional model-comparison visualizations and token-cost analysis appear in Appendices~\ref{app:model-comparison-figs} and~\ref{app:token-cost}.

\subsection{The combined score separates passing and failing trajectories}
\label{sec:complementarity}

After analyzing the main process-quality findings, we validate whether the composite score also captures outcome-relevant behavioral differences beyond the pilot set used for calibration. On the scaled 47-task evaluation set, the combined score reaches AUROC~\citep{auroc} $=0.766$, accuracy $=72.0\%$, F1 $=0.723$, and KS~\citep{ks} $p=0.0017$ for separating passing and failing trajectories. This experiment is a sanity check, not a replacement for pass/fail evaluation: if process quality were unrelated to outcome, the score would be difficult to interpret. The value of \agentlens comes from the mismatch cases. Passing trajectories with low quality scores expose Lucky Passes, while failing trajectories that remain close to known-good processes expose Partial-fail cases that may be recoverable.

The combined score is the only signal with statistically significant pass/fail separation. No individual signal reaches $p<0.05$ (Table~\ref{tab:signals}). This partial agreement with outcome is useful, but incomplete by design: structural alignment and set coverage measure how much of the known-good solution space the agent follows, while coherence and temporal profile measure whether the agent moves through the task in an orderly way. These process signals explain why trajectories with the same binary outcome can receive different quality scores.

\begin{table}[!ht]
\centering
\caption{\textbf{Per-signal AUROC.} No individual signal achieves $p<0.05$; only the combined score significantly separates passing and failing trajectories.}
\label{tab:signals}
\small
\begin{tabular}{@{}lccc@{}}
\toprule
\textbf{Signal} & \textbf{AUROC} & \textbf{Weight} & \textbf{KS $p$} \\
\midrule
Structural alignment ($\Phi_{\mathrm{struct}}$) & 0.710 & 0.20 & 0.196 \\
Set coverage ($\Phi_{\mathrm{cov}}$)            & 0.718 & 0.15 & 0.603 \\
Trajectory coherence ($\Phi_{\mathrm{coh}}$)    & 0.728 & 0.30 & 0.603 \\
Temporal profile ($\Phi_{\mathrm{temp}}$)       & 0.653 & 0.35 & 0.365 \\
\midrule
\textbf{Combined} & \textbf{0.766} & 1.00 & \textbf{0.0017} \\
\bottomrule
\end{tabular}
\vspace{-6pt}
\end{table}

We also compare \agentlens with simpler trajectory-similarity baselines. \agentlens outperforms TF-IDF and dense embedding alignment by 0.065--0.094 AUROC. Individual trajectory matching slightly outperforms \agentlens in AUROC (0.805), but it requires $O(N_{\mathrm{train}})$ comparisons at inference and returns only a scalar similarity to one best-match trace. The merged-PTA representation is more useful for analysis because it localizes divergence, measures branch-level coverage, and attributes waste relative to known-good solution paths. Appendix~\ref{app:baseline-results} reports the full baseline table, and Appendix~\ref{app:score-distributions} shows the score distributions.

Among failing trajectories, 54.9\% are Partial-fail and 45.1\% are Off-track. Partial-fail trajectories remain close enough to known-good processes to exceed the failure-tier threshold, while Off-track trajectories diverge earlier or more completely. This split suggests that roughly half of failures are structurally recoverable: the agent follows a reasonable strategy but makes a localized error.

\subsection{Intent-stage labels are reliable}
\label{sec:rq-labels}

Before using intent-stage labels for scoring, we validate them with a seven-annotator agreement study. Following the MAST protocol~\citep{cemri2025mast}, five human annotators, all software engineers who use coding agents in their daily work, and two LLM annotators labeled 200 deduplicated state-actions sampled across tools and phases. The labels reach Fleiss' $\kappa=0.933$ with 96.0\% raw agreement. Of the 200 states, 192 reached consensus; evaluated against these consensus labels, the \agentlens heuristic achieves 93.8\% accuracy and macro-F1 $=0.933$. The remaining disagreements concentrate on post-implementation \texttt{read\_file} calls, the boundary case targeted by the context-sensitive labeler. Appendix~\ref{app:heuristic-validation} gives the per-stage breakdown.

% ============================================================
%  6  ROBUSTNESS AND ABLATIONS
% ============================================================
\section{Robustness and Ablations}
\label{sec:ablation}

We use the disjoint 278-trajectory pilot set for controlled ablations. The goal is not to re-tune the score, but to test whether the main design choices are necessary: combining structural and behavioral signals, merging multiple passing trajectories into a PTA, and fixing a merge count for scaled evaluation. Full tables are reported in Appendix~\ref{app:ablation-details}.

\paragraph{Signal contribution.}
Removing any one signal reduces AUROC relative to the full score. The largest drops come from removing temporal profile divergence ($-0.037$) and trajectory coherence ($-0.031$), followed by set coverage ($-0.024$) and structural alignment ($-0.016$). This supports the design choice that structural signals and behavioral signals should be fused rather than treated as substitutes. Structural alignment and coverage measure whether the agent visits relevant states; coherence and temporal profile measure whether it moves through them in a plausible problem-solving order. Weight sensitivity is mild: perturbing any single weight by $\pm 0.05$ reduces AUROC by at most 0.006. The selected weights are therefore one well-supported configuration; the conclusions rely on the stability of the resulting score rather than on the exact coefficients.

\paragraph{Merge-count sensitivity.}
The number of passing trajectories merged into the PTA controls a precision--coverage trade-off. With small $k$ (here, $k$ represents the number of passing trajectories used to construct the merged PTA), the PTA is compact and precise but may penalize valid strategies absent from the reference. With larger $k$, the PTA covers more successful strategies but becomes more permissive and can exceed scoring limits on complex tasks. On the pilot set, $k=2$ achieves AUROC $=0.749$ with full task coverage, while $k=5$ achieves AUROC $=0.777$ and covers a broader solution space. At $k\geq 6$, AUROC rises further but only a small subset of easier-to-score task resamples remains, indicating survivorship bias rather than genuine improvement. We therefore use $k=5$ for scaled evaluation.

\paragraph{Merge-order robustness.}
Because PTA construction is incremental, we also test whether merge order materially changes downstream scores.
% On \texttt{astropy\_\_astropy-12907},
On one of the tasks, we evaluate 10 trajectory combinations and all 6 merge permutations for each combination. Trajectory selection explains 64.1\% of the variance, while merge ordering explains 35.9\%; 7 of 10 combinations are fully order-invariant. The remaining order effects occur when an ambiguous exploration prefix can be either merged or branched, but the observed range is bounded and smaller than the effect of which trajectories are selected. Thus, the main source of PTA variation is reference-set choice, not merge ordering.

% ============================================================
%  7  CONCLUSION
% ============================================================
\section{Conclusion}
\label{sec:conclusion}

In this work, we introduced \agentlens, a process-aware framework for evaluating SWE-agent trajectories beyond binary pass/fail outcomes. \agentlens converts raw trajectories into intent-labeled state sequences and merges passing trajectories into task-level PTA references. This lets us score not only whether an agent reached a correct patch, but whether it followed a coherent and low-waste solution process. With this view, \agentlens distinguishes direct solutions, valid alternative paths, weak successful trajectories that pass only after brittle behavior, and failed trajectories that remain close to known-good processes.

Evaluated on 2{,}614 OpenHands trajectories from 60 SWE-bench Verified tasks, \agentlens reveals substantial variation among successful runs. On the 47 PTA-eligible tasks that form \agentlens-Bench, 10.7\% of passing trajectories are Lucky Passes: correct outcomes reached through weak processes. These weak successes follow recurring patterns rather than isolated accidents, with Brute-Force Convergence and Incomplete Implementation accounting for 68.0\% of Lucky Passes. Process-aware scoring also changes model comparison: across all eight evaluated model backends, quality-score rankings differ from pass-rate rankings, and Lucky rates range from 0.5\% to 23.2\%. We release \agentlens-Bench, the \agentlens SDK, and the web interface at \url{https://github.com/agentlens-framework/agentlens} to make process-aware analysis a standard layer in coding-agent evaluation.

\section*{Acknowledgments and Disclosure of Funding}
This work was supported by Microsoft. Priyam Sahoo was a student researcher at Microsoft during this work. Competing interests: Gaurav Mittal, Xiaomin Li, Shengjie Ma, Benjamin Steenhoek, Pingping Lin, and Yu Hu are employees of Microsoft.

\bibliographystyle{plainnat}
\bibliography{references}

@inproceedings{badertdinov2025swe,
  title={{SWE-rebench}: An Automated Pipeline for Task Collection and Decontaminated Evaluation of Software Engineering Agents},
  author={Badertdinov, Ibragim and Golubev, Alexander and Nekrashevich, Maksim and Shevtsov, Anton and Karasik, Simon and Andriushchenko, Andrei and Trofimova, Maria and Litvintseva, Daria and Yangel, Boris},
  booktitle={Advances in Neural Information Processing Systems},
  volume={38},
  year={2025}
}

@inproceedings{bouzenia2025understanding,
  title={Understanding Software Engineering Agents: A Study of Thought-Action-Result Trajectories},
  author={Bouzenia, Islem and Pradel, Michael},
  booktitle={2025 40th IEEE/ACM International Conference on Automated Software Engineering (ASE)},
  pages={2846--2857},
  year={2025}
}

@misc{harbor_atif,
  title        = {{ATIF}: Agent Trajectory Interchange Format},
  author       = {{Harbor Framework}},
  howpublished = {\url{https://docs.harborframework.com/agents/atif}},
  year         = {2026},
  note         = {Accessed 2026-05-05}
}

@article{devin,
        author = {Sana Ansari and Sakshi Kini},
        title = {The World's First {AI} Software Engineer, {DEVIN AI}},
        journal = {International Journal of Innovative Research in Technology},
        year = {2024},
        volume = {11},
        number = {1},
        pages = {1813-1816},
        issn = {2349-6002},
        url = {https://ijirt.org/article?manuscript=165794},
        month = {June},
        }

@article{deshpande2025trail,
  title={{TRAIL}: Trace reasoning and agentic issue localization},
  author={Deshpande, Darshan and Gangal, Varun and Mehta, Hersh and Krishnan, Jitin and Kannappan, Anand and Qian, Rebecca},
  journal={arXiv preprint arXiv:2505.08638},
  year={2025}
}

@article{fawcett2006introduction,
  title={An introduction to {ROC} analysis},
  author={Fawcett, Tom},
  journal={Pattern recognition letters},
  volume={27},
  number={8},
  pages={861--874},
  year={2006},
  publisher={Elsevier}
}

@inproceedings{jain2024livecodebench,
  title={{LiveCodeBench}: Holistic and Contamination Free Evaluation of Large Language Models for Code},
  author={Jain, Naman and Han, King and Gu, Alex and Li, Wen-Ding and Yan, Fanjia and Zhang, Tianjun and Wang, Sida and Solar-Lezama, Armando and Sen, Koushik and Stoica, Ion},
  booktitle={The Thirteenth International Conference on Learning Representations},
  year={2025}
}

@article{jain2025r2e,
  title={{R2E-Gym}: Procedural environments and hybrid verifiers for scaling open-weights {SWE} agents},
  author={Jain, Naman and Singh, Jaskirat and Shetty, Manish and Zheng, Liang and Sen, Koushik and Stoica, Ion},
  journal={arXiv preprint arXiv:2504.07164},
  year={2025}
}

@article{ko2006exploratory,
  title={An exploratory study of how developers seek, relate, and collect relevant information during software maintenance tasks},
  author={Ko, Amy J and Myers, Brad A and Coblenz, Michael J and Aung, Htet Htet},
  journal={IEEE Transactions on software engineering},
  volume={32},
  number={12},
  pages={971--987},
  year={2006},
  publisher={IEEE}
}

@article{liu2026process,
  title={Process-Centric Analysis of Agentic Software Systems},
  author={Liu, Shuyang and Chen, Yang and Krishna, Rahul and Sinha, Saurabh and Ganhotra, Jatin and Jabbarvand, Reyhaneh},
  journal={Proceedings of the ACM on Programming Languages},
  volume={10},
  number={OOPSLA1},
  pages={1961--1988},
  year={2026},
  publisher={ACM New York, NY, USA}
}

@article{ma2024agentboard,
  title={{AgentBoard}: An analytical evaluation board of multi-turn {LLM} agents},
  author={Ma, Chang and Zhang, Junlei and Zhu, Zhihao and Yang, Cheng and Yang, Yujiu and Jin, Yaohui and Lan, Zhenzhong and Kong, Lingpeng and He, Junxian},
  journal={Advances in neural information processing systems},
  volume={37},
  pages={74325--74362},
  year={2024}
}

@article{majgaonkar2025understanding,
  title={Understanding Code Agent Behaviour: An Empirical Study of Success and Failure Trajectories},
  author={Majgaonkar, Oorja and Fei, Zhiwei and Li, Xiang and Sarro, Federica and Ye, He},
  journal={arXiv preprint arXiv:2511.00197},
  year={2025}
}

@incollection{oncina1992inferring,
  title={Inferring regular languages in polynomial update time},
  author={Oncina, Jos{\'e} and Garc{\'i}a, Pedro},
  booktitle={Pattern Recognition and Image Analysis},
  series={Series in Machine Perception and Artificial Intelligence},
  volume={1},
  pages={49--61},
  year={1992},
  publisher={World Scientific}
}

@article{uesato2022solving,
  title={Solving math word problems with process- and outcome-based feedback},
  author={Uesato, Jonathan and Kushman, Nate and Kumar, Ramana and Song, Francis and Siegel, Noah and Wang, Lisa and Creswell, Antonia and Irving, Geoffrey and Higgins, Irina},
  journal={arXiv preprint arXiv:2211.14275},
  year={2022}
}

@inproceedings{wei2025swe,
  title={{SWE-RL}: Advancing {LLM} Reasoning via Reinforcement Learning on Open Software Evolution},
  author={Wei, Yuxiang and Duchenne, Olivier and Copet, Jade and Carbonneaux, Quentin and Zhang, Lingming and Fried, Daniel and Synnaeve, Gabriel and Singh, Rishabh and Wang, Sida I},
  booktitle={Advances in Neural Information Processing Systems},
  volume={38},
  pages={78500--78525},
  year={2025}
}

@article{xia2024agentless,
  title={Demystifying {LLM}-Based Software Engineering Agents},
  author={Xia, Chunqiu Steven and Deng, Yinlin and Dunn, Soren and Zhang, Lingming},
  journal={Proceedings of the ACM on Software Engineering},
  volume={2},
  number={FSE},
  pages={801--824},
  year={2025}
}

@article{auroc,
  title = {The use of the area under the {ROC} curve in the evaluation of machine learning algorithms},
  journal = {Pattern Recognition},
  volume = {30},
  number = {7},
  pages = {1145-1159},
  year = {1997},
  issn = {0031-3203},
  doi = {https://doi.org/10.1016/S0031-3203(96)00142-2},
  url = {https://www.sciencedirect.com/science/article/pii/S0031320396001422},
  author = {Andrew P. Bradley}
}

@article{ks,
  ISSN = {01621459, 1537274X},
  URL = {http://www.jstor.org/stable/2280095},
  author = {Frank J. Massey},
  journal = {Journal of the American Statistical Association},
  number = {253},
  pages = {68--78},
  publisher = {[American Statistical Association, Taylor & Francis, Ltd.]},
  title = {The {Kolmogorov-Smirnov} Test for Goodness of Fit},
  urldate = {2026-05-06},
  volume = {46},
  year = {1951}
}

@article{youden1950index,
    author = {Youden, W. J.},
    title = {Index for rating diagnostic tests},
    journal = {Cancer},
    volume = {3},
    number = {1},
    pages = {32-35},
    doi = {https://doi.org/10.1002/1097-0142(1950)3:1<32::AID-CNCR2820030106>3.0.CO;2-3},
    url = {https://acsjournals.onlinelibrary.wiley.com/doi/abs/10.1002/1097-0142%281950%293%3A1%3C32%3A%3AAID-CNCR2820030106%3E3.0.CO%3B2-3},
    eprint = {https://acsjournals.onlinelibrary.wiley.com/doi/pdf/10.1002/1097-0142%281950%293%3A1%3C32%3A%3AAID-CNCR2820030106%3E3.0.CO%3B2-3},
    year = {1950}
}

@inproceedings{zhang2024autocoderover,
    author = {Zhang, Yuntong and Ruan, Haifeng and Fan, Zhiyu and Roychoudhury, Abhik},
    title = {{AutoCodeRover}: Autonomous Program Improvement},
    year = {2024},
    isbn = {9798400706127},
    publisher = {Association for Computing Machinery},
    address = {New York, NY, USA},
    url = {https://doi.org/10.1145/3650212.3680384},
    doi = {10.1145/3650212.3680384},
    booktitle = {Proceedings of the 33rd ACM SIGSOFT International Symposium on Software Testing and Analysis},
    pages = {1592--1604},
    numpages = {13},
    location = {Vienna, Austria},
    series = {ISSTA 2024}
}

@inproceedings{zhu2025abc,
  title={Establishing Best Practices in Building Rigorous Agentic Benchmarks},
  author={Zhu, Yuxuan and Jin, Tengjun and Pruksachatkun, Yada and Zhang, Andy and Liu, Shu and Cui, Sasha and Kapoor, Sayash and Longpre, Shayne and Meng, Kevin and Weiss, Rebecca and Barez, Fazl and Gupta, Rahul and Dhamala, Jwala and Merizian, Jacob and Giulianelli, Mario and Coppock, Harry and Ududec, Cozmin and Kellermann, Antony and Sekhon, Jasjeet and Steinhardt, Jacob and Schwettmann, Sarah and Narayanan, Arvind and Zaharia, Matei A and Stoica, Ion and Liang, Percy and Kang, Daniel},
  booktitle={Advances in Neural Information Processing Systems},
  volume={38},
  year={2025}
}

@inproceedings{zhuge2024agentasjudge,
  title={{Agent-as-a-Judge}: Evaluate Agents with Agents},
  author={Zhuge, Mingchen and Zhao, Changsheng and Ashley, Dylan R. and Wang, Wenyi and Khizbullin, Dmitrii and Xiong, Yunyang and Liu, Zechun and Chang, Ernie and Krishnamoorthi, Raghuraman and Tian, Yuandong and Shi, Yangyang and Chandra, Vikas and Schmidhuber, J{\"u}rgen},
  booktitle={Proceedings of the 42nd International Conference on Machine Learning},
  series={Proceedings of Machine Learning Research},
  volume={267},
  pages={80569--80611},
  publisher={PMLR},
  year={2025}
}

@inproceedings{zhuo2024bigcodebench,
  title={{BigCodeBench}: Benchmarking Code Generation with Diverse Function Calls and Complex Instructions},
  author={Zhuo, Terry Yue and Vu, Minh Chien and Chim, Jenny and Hu, Han and Yu, Wenhao and Widyasari, Ratnadira and Yusuf, Imam Nur Bani and Zhan, Haolan and He, Junda and Paul, Indraneil and others},
  booktitle={The Thirteenth International Conference on Learning Representations},
  year={2025}
}

@article{alaboudi2021exploratory,
  author = {Alaboudi, Abdulaziz and LaToza, Thomas D.},
  title = {An Exploratory Study of Debugging Episodes},
  journal = {arXiv preprint arXiv:2105.02162},
  year = {2021}
}

@misc{brunsfeld2024treesitter,
  author = {Brunsfeld, Max and others},
  title = {tree-sitter/tree-sitter: v0.23.0},
  howpublished = {Software, \url{https://github.com/tree-sitter/tree-sitter}},
  year = {2024}
}

@inproceedings{cemri2025mast,
  title={Why Do Multi-Agent {LLM} Systems Fail?},
  author={Cemri, Mert and Pan, Melissa Z and Yang, Shuyi and Agrawal, Lakshya A and Chopra, Bhavya and Tiwari, Rishabh and Keutzer, Kurt and Parameswaran, Aditya and Klein, Dan and Ramchandran, Kannan and others},
  booktitle={Advances in Neural Information Processing Systems},
  volume={38},
  year={2025}
}

@article{deepmind2024gemini,
  title={Gemini 2.5: Pushing the frontier with advanced reasoning, multimodality, long context, and next generation agentic capabilities},
  author={Comanici, Gheorghe and Bieber, Eric and Schaekermann, Mike and Pasupat, Ice and Sachdeva, Noveen and Dhillon, Inderjit and Blistein, Marcel and Ram, Ori and Zhang, Dan and Rosen, Evan and others},
  journal={arXiv preprint arXiv:2507.06261},
  year={2025}
}

@inproceedings{jimenez2024swe,
  author = {Jimenez, Carlos E. and Yang, John and Wettig, Alexander and Yao, Shunyu and Pei, Kexin and Press, Ofir and Narasimhan, Karthik},
  title = {{SWE-bench}: Can Language Models Resolve Real-World {GitHub} Issues?},
  booktitle = {The Twelfth International Conference on Learning Representations},
  year = {2024}
}

@inproceedings{lightman2023verify,
  title={Let's verify step by step},
  author={Lightman, Hunter and Kosaraju, Vineet and Burda, Yuri and Edwards, Harrison and Baker, Bowen and Lee, Teddy and Leike, Jan and Schulman, John and Sutskever, Ilya and Cobbe, Karl},
  booktitle={The Twelfth International Conference on Learning Representations},
  year={2024}
}

@article{lin1991divergence,
  author = {Lin, J.},
  title = {Divergence Measures Based on the {Shannon} Entropy},
  journal = {IEEE Transactions on Information Theory},
  volume = {37},
  number = {1},
  pages = {145--151},
  year = {1991}
}

@inproceedings{liu2024agentbench,
  title={{AgentBench}: Evaluating {LLMs} as Agents},
  author={Liu, Xiao and Yu, Hao and Zhang, Hanchen and Xu, Yifan and Lei, Xuanyu and Lai, Hanyu and Gu, Yu and Ding, Hangliang and Men, Kaiwen and Yang, Kejuan and others},
  booktitle={The Twelfth International Conference on Learning Representations},
  year={2024}
}

@inproceedings{miserendino2025swelancer,
  title={{SWE-Lancer}: Can Frontier {LLMs} Earn \$1 Million from Real-World Freelance Software Engineering?},
  author={Miserendino, Samuel and Wang, Michele and Patwardhan, Tejal and Heidecke, Johannes},
  booktitle={Proceedings of the 42nd International Conference on Machine Learning},
  series={Proceedings of Machine Learning Research},
  volume={267},
  pages={44412--44450},
  publisher={PMLR},
  year={2025}
}

@misc{openai2024swebenchverified,
  title={Introducing {SWE-bench} Verified},
  author={Chowdhury, Neil and Aung, James and Shern, Chan Jun and Jaffe, Oliver and Sherburn, Dane and Starace, Giulio and Mays, Evan and Dias, Rachel and Aljubeh, Marwan and Glaese, Mia and others},
  howpublished={OpenAI blog, \url{https://openai.com/index/introducing-swe-bench-verified/}},
  year={2024}
}

@inproceedings{pan2025swe,
  author = {Pan, Jiayi and Wang, Xingyao and Neubig, Graham and Jaitly, Navdeep and Ji, Heng and Suhr, Alane and Zhang, Yizhe},
  title = {Training Software Engineering Agents and Verifiers with {SWE-Gym}},
  booktitle = {Proceedings of the 42nd International Conference on Machine Learning},
  series = {Proceedings of Machine Learning Research},
  volume = {267},
  pages = {47717--47737},
  publisher = {PMLR},
  year = {2025}
}

@inproceedings{pan2025webshepherd,
  title={{Web-Shepherd}: Advancing {PRMs} for Reinforcing Web Agents},
  author={Chae, Hyungjoo and Kim, Seonghwan and Cho, Junhee and Kim, Seungone and Moon, Seungjun and Hwangbo, Gyeom and Lim, Dongha and Kim, Minjin and Hwang, Yeonjun and Gwak, Minju and others},
  booktitle={Advances in Neural Information Processing Systems},
  volume={38},
  pages={63314--63356},
  year={2025}
}

@article{shum2025swerm,
  title={{SWE-RM}: Execution-free Feedback For Software Engineering Agents},
  author={Shum, KaShun and Hui, Binyuan and Chen, Jiawei and Zhang, Lei and W., X. and Yang, Jiaxi and Huang, Yuzhen and Lin, Junyang and He, Junxian},
  journal={arXiv preprint arXiv:2512.21919},
  year={2025}
}

@inproceedings{stanford2025terminalbench,
  title={{Terminal-Bench}: Benchmarking Agents on Hard, Realistic Tasks in Command Line Interfaces},
  author={Merrill, Mike A and Shaw, Alexander G and Carlini, Nicholas and Li, Boxuan and Raj, Harsh and Bercovich, Ivan and Shi, Lin and Shin, Jeong Yeon and Walshe, Thomas and Buchanan, E Kelly and others},
  booktitle={The Fourteenth International Conference on Learning Representations},
  year={2026}
}

@inproceedings{wang2024mathshepherd,
  title={{Math-Shepherd}: Verify and reinforce {LLMs} step-by-step without human annotations},
  author={Wang, Peiyi and Li, Lei and Shao, Zhihong and Xu, Runxin and Dai, Damai and Li, Yifei and Chen, Deli and Wu, Yu and Sui, Zhifang},
  booktitle={Proceedings of the 62nd Annual Meeting of the Association for Computational Linguistics (Volume 1: Long Papers)},
  pages={9426--9439},
  year={2024}
}

@inproceedings{zhang2025swebenchlive,
  title = {{SWE-bench Goes Live!}},
  author = {Zhang, Linghao and He, Shilin and Zhang, Chaoyun and Kang, Yu and Li, Bowen and Xie, Chengxing and Wang, Junhao and Wang, Maoquan and Huang, Yufan and Fu, Shengyu and others},
  booktitle = {Advances in Neural Information Processing Systems},
  volume = {38},
  year = {2025}
}

@inproceedings{wang2025openhands,
  title={{OpenHands}: An Open Platform for {AI} Software Developers as Generalist Agents},
  author={Wang, Xingyao and Li, Boxuan and Song, Yufan and Xu, Frank F and Tang, Xiangru and Zhuge, Mingchen and Pan, Jiayi and Song, Yueqi and Li, Bowen and Singh, Jaskirat and others},
  booktitle={The Thirteenth International Conference on Learning Representations},
  year={2025}
}

@inproceedings{wang2025multiswe,
  title={{Multi-SWE-bench}: A Multilingual Benchmark for Issue Resolving},
  author={Zan, Daoguang and Huang, Zhirong and Liu, Wei and Chen, Hanwu and Xin, Shulin and Zhang, Linhao and Liu, Qi and Li, Aoyan and Chen, Lu and Zhong, Xiaojian and others},
  booktitle={Advances in Neural Information Processing Systems},
  volume={38},
  year={2025}
}

@inproceedings{xie2024osworld,
  author = {Xie, Tianbao and Zhang, Danyang and Chen, Jixuan and Li, Xiaochuan and Zhao, Siheng and Cao, Ruisheng and Hua, Toh Jing and Cheng, Zhoujun and Shin, Dongchan and Lei, Fangyu and Liu, Yitao and Xu, Yiheng and Zhou, Shuyan and Savarese, Silvio and Xiong, Caiming and Zhong, Victor and Yu, Tao},
  title = {{OSWorld}: Benchmarking Multimodal Agents for Open-Ended Tasks in Real Computer Environments},
  booktitle = {Advances in Neural Information Processing Systems},
  volume = {37},
  pages = {52040--52094},
  year = {2024}
}

@inproceedings{yang2024swe,
    author = {Yang, John and Jimenez, Carlos and Wettig, Alexander and Lieret, Kilian and Yao, Shunyu and Narasimhan, Karthik and Press, Ofir},
    booktitle = {Advances in Neural Information Processing Systems},
    doi = {10.52202/079017-1601},
    editor = {A. Globerson and L. Mackey and D. Belgrave and A. Fan and U. Paquet and J. Tomczak and C. Zhang},
    pages = {50528--50652},
    publisher = {Curran Associates, Inc.},
    title = {{SWE-agent}: Agent-Computer Interfaces Enable Automated Software Engineering},
    url = {https://proceedings.neurips.cc/paper_files/paper/2024/file/5a7c947568c1b1328ccc5230172e1e7c-Paper-Conference.pdf},
    volume = {37},
    year = {2024}
}

@inproceedings{yang2025swesmith,
  title={{SWE-smith}: Scaling Data for Software Engineering Agents},
  author={Yang, John and Lieret, Kilian and Jimenez, Carlos E. and Wettig, Alexander and Khandpur, Kabir and Zhang, Yanzhe and Hui, Binyuan and Press, Ofir and Schmidt, Ludwig and Yang, Diyi},
  booktitle={Advances in Neural Information Processing Systems},
  volume={38},
  year={2025}
}

@inproceedings{zhang2020bertscore,
  title={{BERTScore}: Evaluating Text Generation with {BERT}},
  author={Zhang, Tianyi and Kishore, Varsha and Wu, Felix and Weinberger, Kilian Q. and Artzi, Yoav},
  booktitle={International Conference on Learning Representations},
  year={2020}
}

@inproceedings{zheng2025processbench,
    title = "{P}rocess{B}ench: Identifying Process Errors in Mathematical Reasoning",
    author = "Zheng, Chujie  and
      Zhang, Zhenru  and
      Zhang, Beichen  and
      Lin, Runji  and
      Lu, Keming  and
      Yu, Bowen  and
      Liu, Dayiheng  and
      Zhou, Jingren  and
      Lin, Junyang",
    editor = "Che, Wanxiang  and
      Nabende, Joyce  and
      Shutova, Ekaterina  and
      Pilehvar, Mohammad Taher",
    booktitle = "Proceedings of the 63rd Annual Meeting of the Association for Computational Linguistics (Volume 1: Long Papers)",
    month = jul,
    year = "2025",
    address = "Vienna, Austria",
    publisher = "Association for Computational Linguistics",
    url = "https://aclanthology.org/2025.acl-long.50/",
    doi = "10.18653/v1/2025.acl-long.50",
    pages = "1009--1024",
    ISBN = "979-8-89176-251-0"
}

@inproceedings{zhou2024webarena,
  title={{WebArena}: A Realistic Web Environment for Building Autonomous Agents},
  author={Zhou, Shuyan and Xu, Frank F and Zhu, Hao and Zhou, Xuhui and Lo, Robert and Sridhar, Abishek and Cheng, Xianyi and Ou, Tianyue and Bisk, Yonatan and Fried, Daniel and others},
  booktitle={The Twelfth International Conference on Learning Representations},
  year={2024}
}

@misc{PTA_sharma2026,
      title={Learning Correct Behavior from Examples: Validating Sequential Execution in Autonomous Agents}, 
      author={Reshabh K Sharma and Gaurav Mittal and Yu Hu},
      year={2026},
      eprint={2605.03159},
      archivePrefix={arXiv},
      primaryClass={cs.AI},
      url={https://arxiv.org/abs/2605.03159}, 
}

\newpage

\appendix
% ============================================================
%  APPENDIX CONTENTS
% ============================================================
\makeatletter
\let\agentlensorigsection\section
\let\agentlensorigsubsection\subsection
\RenewDocumentCommand{\section}{s o m}{%
  \IfBooleanTF{#1}
    {\agentlensorigsection*{#3}}%
    {%
      \IfNoValueTF{#2}{\agentlensorigsection{#3}}{\agentlensorigsection[#2]{#3}}%
      \addcontentsline{atoc}{section}{\protect\numberline{\thesection}\IfNoValueTF{#2}{#3}{#2}}%
    }%
}
\RenewDocumentCommand{\subsection}{s o m}{%
  \IfBooleanTF{#1}
    {\agentlensorigsubsection*{#3}}%
    {%
      \IfNoValueTF{#2}{\agentlensorigsubsection{#3}}{\agentlensorigsubsection[#2]{#3}}%
      \addcontentsline{atoc}{subsection}{\protect\numberline{\thesubsection}\IfNoValueTF{#2}{#3}{#2}}%
    }%
}
\newcommand{\agentlensappendixtocentry}[4]{%
  \par
  \addvspace{#1}%
  \noindent
  \begingroup
    \parindent=0pt
    \rightskip=0pt plus 1fil
    \parfillskip=0pt
    \leavevmode
    \hspace*{#2}{#3}%
    \nobreak\leaders\hbox to 0.65em{\hss.\hss}\hfill
    \nobreak\hb@xt@ 2.5em{\hfil\color{black}#4}\par
  \endgroup
}
\newcommand{\agentlensappendixsectiontoc}[2]{%
  \begingroup
  \def\numberline##1{\makebox[1.8em][l]{##1}}%
  \agentlensappendixtocentry{0.12em}{0pt}{\bfseries #1}{#2}%
  \endgroup
}
\newcommand{\agentlensappendixsubsectiontoc}[2]{%
  \begingroup
  \def\numberline##1{\makebox[3.2em][l]{##1}}%
  \agentlensappendixtocentry{-0.10em}{2.2em}{#1}{#2}%
  \endgroup
}
\makeatother

% Tighter appendix float spacing keeps figures close without leaving large gaps.
\setlength{\textfloatsep}{8pt plus 2pt minus 2pt}
\setlength{\floatsep}{6pt plus 2pt minus 2pt}
\setlength{\intextsep}{6pt plus 2pt minus 2pt}

\section*{Appendix}
% \vspace{0.5em}

{\bfseries Table of Contents\par}
\vspace{-1.3em}
\noindent\rule{\linewidth}{0.5pt}

{
% \small
\setlength{\parskip}{2.5pt}
\setcounter{tocdepth}{2}
\makeatletter
\begingroup
\let\l@section\agentlensappendixsectiontoc
\let\l@subsection\agentlensappendixsubsectiontoc
\@starttoc{atoc}
\endgroup
\makeatother
}

\newpage

% ============================================================
%  APPENDIX A: LIMITATIONS, FUTURE DIRECTIONS, POSITIONING
% ============================================================

\section{Limitations, Future Directions, and Positioning}
\label{app:limitations-future-derections-positioning}

\subsection{Limitations}
\label{app:limitations}

\agentlens is designed as a post-hoc analysis framework for SWE-agent trajectories, and our evaluation focuses on OpenHands-style coding-agent traces on SWE-bench Verified tasks. This scope lets us study process quality in a controlled setting with task-level PTA references and reproducible scoring. Other agent scaffolds can be connected through ATIF conversion or trace-format adapters without changing the core scoring pipeline, but our empirical results are validated only on OpenHands trajectories: scaffold-specific tool names and command conventions may require updates to the tool registry and command-level rules before scores are comparable across scaffolds. Similarly, the fixed score weights used in this paper are chosen to keep the benchmark deterministic and comparable across models; future deployments may tune these weights for domain-specific priorities such as verification discipline, exploration cost, or repair efficiency.

\subsection{Future Directions}
\label{app:future-directions}

\agentlens opens several directions for process-aware agent research. Its tiers already provide a finer data-selection signal than pass/fail: they rank successful demonstrations by process quality and identify failed trajectories whose prefixes follow a known-good process (the 54.9\% of failures labeled Partial-fail). Measuring the downstream benefit requires a controlled comparison that fine-tunes the same model on equal-sized sets of quality-selected and randomly selected passing trajectories and evaluates both on held-out tasks. First, quality scores can be used as dense reward signals for reinforcement learning, encouraging agents not only to reach correct patches but to follow coherent, low-waste solution processes~\citep{wei2025swe}. Second, PTA references can support curriculum construction: training pools can be organized by process quality, divergence type, or recoverability rather than by final outcome alone. Third, longitudinal process-quality tracking can compare model versions over time, revealing whether higher pass rates come from cleaner reasoning, broader exploration, or increased retry behavior. Finally, the same trajectory-analysis view can extend beyond software repair to web navigation~\citep{zhou2024webarena} and computer-use agents~\citep{xie2024osworld}, where success alone often hides large differences in action efficiency and recoverability.

\subsection{Broader Impacts}
\label{app:broader-impacts}

\agentlens is intended to improve the evaluation and analysis of software engineering agents. By exposing brittle successful trajectories, recoverable failures, and wasteful solution processes, it can help practitioners make safer deployment decisions, select cleaner training demonstrations, and diagnose weaknesses that are hidden by pass/fail benchmarks. This can improve benchmark incentives by rewarding agents that solve tasks coherently rather than through excessive retry or accidental success.

At the same time, process scores should be treated as complementary diagnostics rather than replacements for functional correctness, security review, or human judgment. A trajectory that receives a high \agentlens score may still contain an incorrect or insecure patch if the underlying tests or references are incomplete. We therefore position \agentlens as an additional layer for responsible evaluation, not as an automatic approval mechanism for deploying code changes.

\subsection{Extended Related Work and Positioning}
\label{app:extended-related}

Table~\ref{tab:dataset_comparison} compares \agentlens-Bench against related SWE-agent trajectory collections. Table~\ref{tab:positioning} compares \agentlens against related evaluation frameworks.

\begin{table}[!ht]
\centering
\caption{\agentlens-Bench versus related trajectory collections. Dashes indicate the feature is absent; ``Partial'' indicates a related but incomplete feature.}
\label{tab:dataset_comparison}
\small
\resizebox{\linewidth}{!}{%
\begin{tabular}{@{}lccccc@{}}
\toprule
\textbf{Feature} & \textbf{SWE-Gym} & \textbf{R2E-Gym} & \textbf{OpenHands} & \textbf{Graphectory} & \textbf{\agentlens-Bench} \\
\midrule
Trajectories            & $\sim$6K   & $\sim$3.3K & $\sim$300 & $\sim$4K  & \textbf{1{,}815} \\
Models covered          & 3          & 1         & 5         & 4         & \textbf{8} \\
Pass/fail labels        & \checkmark & \checkmark & \checkmark & \checkmark & \checkmark \\
Quality score (0--100)  & ---        & ---       & ---       & ---       & \checkmark \\
Multi-dim.\ metrics     & ---        & ---       & ---       & \checkmark & \checkmark \\
Ground-truth PTAs       & ---        & ---       & ---       & ---       & \checkmark \\
Waste annotations       & ---        & ---       & ---       & Partial   & \checkmark \\
Divergence localization & ---        & ---       & ---       & ---       & \checkmark \\
Quality tier labels     & ---        & ---       & ---       & ---       & \checkmark \\
Curation support        & Outcome    & Outcome   & ---       & ---       & \checkmark \\
No LLM calls needed     & ---        & ---       & ---       & \checkmark & \checkmark \\
\bottomrule
\end{tabular}
}
\end{table}

\begin{table}[!ht]
\centering
\caption{Framework comparison. \agentlens is the only system that combines SWE-specific trajectory analysis, multi-dimensional scoring, validated phase labels, and structured inefficiency attribution.}
\label{tab:positioning}
\small
\begin{tabular}{@{}lccccc@{}}
\toprule
& \textbf{Domain} & \textbf{Granularity} & \textbf{Multi-dim.} & \textbf{Inefficiency} & \textbf{Label $\kappa$} \\
\midrule
AgentBoard          & Multi (9)    & Sub-goal    & \checkmark & ---     & --- \\
Graphectory         & SWE          & Trace       & \checkmark & ---     & --- \\
MAST                & Multi-agent  & Trace       & \checkmark & ---     & 0.88 \\
TRAIL               & Multi (3)    & Step        & \checkmark & Partial & --- \\
Web-Shepherd        & Web          & Step        & scalar     & ---     & --- \\
SWE-RM              & SWE          & Trajectory  & scalar     & ---     & --- \\
ABC                 & SWE          & Task        & checklist  & ---     & --- \\
\textbf{\agentlens} & \textbf{SWE} & \textbf{Trajectory} & \checkmark & \checkmark & 0.933 \\
\bottomrule
\end{tabular}
\end{table}

% ============================================================
%  APPENDIX B: METHOD DETAILS
% ============================================================

\section{Additional Method Details}
\label{app:method-details}

Section~\ref{sec:method} describes the \agentlens pipeline at the level needed to follow the main argument. This appendix provides the reproducibility-level detail: the full tool registry that drives first-pass label assignment, the state equivalence engine that allows PTA construction to work across heterogeneous agent tool sets, the intent-stage classification decision tree, the formal definitions and worked examples for all four scoring signals, and the planned web interface through which practitioners will be able to inspect the annotations.

The material is organized to mirror the pipeline order. We begin with the tool registry (B.1), which provides the raw mapping from agent tool calls to default intent-stage hints. The equivalence engine (B.2) explains how states from different agents are recognized as covering the same ground-truth action during PTA merging. The intent-stage labeling flow (B.3) gives the full seven-rule priority cascade, including the context-sensitive rules that resolve terminal-command ambiguity. The scoring details (B.4) present the formal definitions, equations, and design rationale for all four signals and the combined score.
Finally, worked examples of all four scoring signals are provided in Appendix~\ref{app:coherence-example}.

\subsection{Tool Registry}
\label{app:tool-registry}

The tool registry provides the first-pass mapping from raw tool calls to intent-stage hints and comparison strategies. These hints are intentionally treated as defaults rather than final labels: the context-sensitive labeler can override them using trajectory history, edited-file state, and command semantics. Table~\ref{tab:registry} shows the abbreviated registry used in the experiments.

\begin{table}[!ht]
\centering
\caption{Tool registry (abbreviated). Each tool maps to a category, a default stage hint, and a comparison strategy.}
\label{tab:registry}
\small
\begin{tabular}{@{}llll@{}}
\toprule
\textbf{Category} & \textbf{Examples} & \textbf{Default Stage} & \textbf{Comparison} \\
\midrule
read       & read\_file, view\_file          & E / V (context) & file path \\
edit       & replace\_string\_in\_file       & I / V (context) & file path \\
search     & grep\_search, semantic\_search  & E               & query \\
execute    & run\_in\_terminal               & E / V (context) & command \\
validation & get\_errors, test\_failure      & V               & identity \\
\bottomrule
\end{tabular}
\end{table}

\subsection{State Equivalence Engine}
\label{app:equivalence}

The equivalence engine determines whether two states from different agents represent the same action. Different agents use different tool names (\texttt{grep} versus \texttt{rg}), read overlapping but not identical file regions, and format terminal arguments differently. The engine applies a confidence-weighted cascade in priority order:

\begin{enumerate}[leftmargin=1.5em, topsep=3pt, itemsep=2pt]
  \item \textbf{Exact content hash} (confidence 1.0): states with identical MD5 hashes are equivalent.
  \item \textbf{File-scope matching} via tree-sitter~\citep{brunsfeld2024treesitter} (confidence 0.90): states targeting the same AST-level scope (function, class, module) are equivalent.
  \item \textbf{Line-range overlap} of at least 30\% (confidence 0.80--0.95): states reading or editing overlapping regions of the same file are equivalent, with confidence scaled by overlap fraction.
  \item \textbf{Semantic terminal grouping} (confidence 0.70--0.85): terminal commands in the same functional group (e.g., \texttt{grep}, \texttt{rg}, \texttt{ag} are all ``search'' commands) with Jaccard token similarity above 0.5 are treated as equivalent.
\end{enumerate}

An optional LLM fallback, which queries a language model for ambiguous cases, was disabled throughout for reproducibility. The cascade is evaluated in order; the first match determines the equivalence decision and its confidence.

\subsection{Intent-stage Labeling Flow}
\label{app:intent-flow}

The intent-stage labeler is the component that most directly affects the behavioral signals. If labels are wrong, coherence and temporal profile become unreliable. The main challenge is terminal commands: \texttt{grep}, \texttt{cat}, and \texttt{ls} can serve exploratory or verificatory purposes depending on where they appear in the trajectory. A naive tool-identity approach labels all terminal commands as a single stage, which causes 70--80\% of PTA states to collapse into one category and renders the temporal signal uninformative.

Figure~\ref{fig:intent-flowchart} shows the full seven-rule priority cascade. Rules 1 through 4 assign fixed stages based on tool type (search tools receive E, validation tools receive V, source edits receive I, orchestration tools receive O). Rules 5 through 7 are context-sensitive: the same tool can map to different stages depending on whether a source-file edit has already occurred, whether the target file is a test file, or whether it was previously modified. The terminal-command sub-tree (Rule 7) disambiguates five command categories by pattern matching. The critical design decision is Rule 5: file-inspection commands (\texttt{grep}, \texttt{cat}, \texttt{ls}, \texttt{git log}, \texttt{find}) receive Exploration regardless of when they appear, because these commands read information from the codebase and do not verify the correctness of any prior edit.

\begin{figure}[h]
  \centering
  \includegraphics[width=\linewidth]{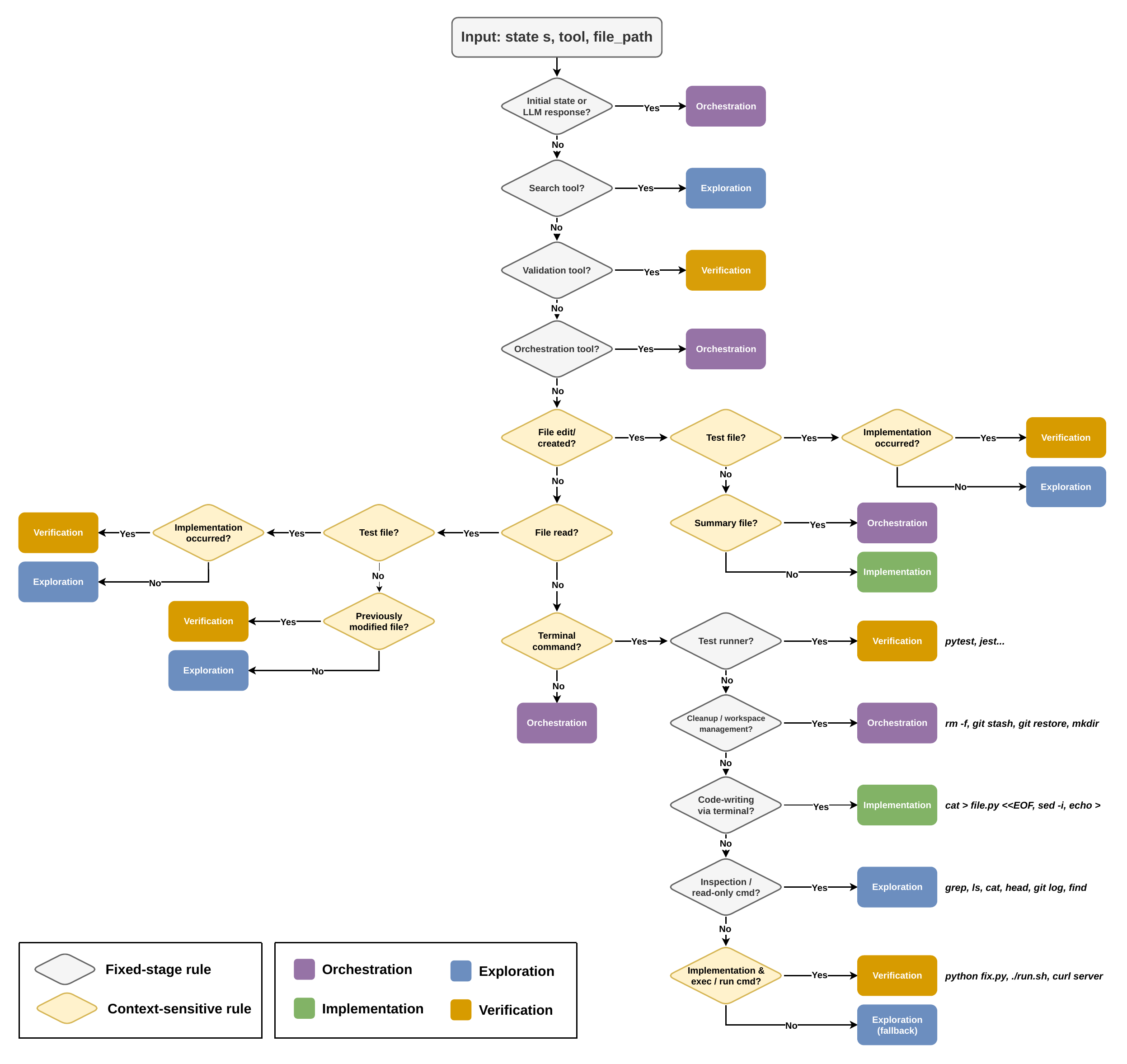}
  \caption{\textbf{Intent-stage classification decision tree.} Each agent action is classified into one of four intent stages (Exploration, Implementation, Verification, Orchestration) via a priority cascade of seven rules. Rules 1 to 4 (gray diamonds) assign fixed stages based on tool type. Rules 5 to 7 (yellow diamonds) are context-sensitive: the same tool can map to different stages depending on whether a source-file edit has already occurred, whether the target file is a test file, or whether it was previously modified. The terminal-command sub-tree (Rule 7) further disambiguates five command categories by pattern matching.}
  \label{fig:intent-flowchart}
\end{figure}

\subsection{Scoring Details}
\label{app:scoring}

Given a candidate trace $\tau_c$ and a ground-truth PTA $\mathcal{G}$, \agentlens computes four signals.

\paragraph{Structural alignment ($\Phi_{\mathrm{struct}}$).}
The candidate's state sequence is aligned against the best-matching PTA path via greedy forward scan (ordered recall) and a greedy one-to-one assignment (unordered precision). Their harmonic mean is the structural F1. This signal measures whether the agent visited the right states in roughly the right order.

\paragraph{Trajectory coherence ($\Phi_{\mathrm{coh}}$).}
The intent-stage sequence is compressed into a workflow fingerprint. Consecutive stage pairs are classified as pivots (forward progress, e.g.\ E$\rightarrow$I), backtracks (regression, e.g.\ V$\rightarrow$E), deepenings (same phase continues), or confirmations (transition to O). The coherence score combines a forward-progress ratio with a blind-retry penalty:
\begin{equation}
  \Phi_{\mathrm{coh}}(\tau) =
  \frac{|\mathrm{pivots}|+|\mathrm{confirms}|}
       {|\mathrm{pivots}|+|\mathrm{confirms}|+|\mathrm{backtracks}|+\epsilon}
  \times
  \left(1-\frac{r}{|T|}\right),
\end{equation}
where $r$ is the blind-retry count and $|T|$ is the total transition count. A clean E$\rightarrow$E$\rightarrow$I$\rightarrow$V trajectory scores 1.0; a trajectory with three regression cycles and a blind-retry cluster scores about 0.51. A worked example of the coherence calculation is provided in Appendix~\ref{app:coherence-example}.

\paragraph{Temporal profile divergence ($\Phi_{\mathrm{temp}}$).}
The trajectory is divided into three equal segments. In each segment the stage distribution is computed with Laplace smoothing ($\alpha=0.01$) and compared against the PTA distribution via Jensen-Shannon divergence~\citep{lin1991divergence}:
\begin{equation}
  \Phi_{\mathrm{temp}}(\tau_c,\mathcal{G}) =
  1-\frac{1}{3}\sum_{k=1}^{3}
  \mathrm{JSD}\!\left(P^{(k)}_{\tau_c}\,\|\,P^{(k)}_{\mathcal{G}}\right).
\end{equation}
This measures whether cognitive phases occurred in the expected temporal order.

\paragraph{Set coverage ($\Phi_{\mathrm{cov}}$).}
The fraction of PTA states across all paths matched by any state in the candidate trajectory, computed via a greedy one-to-one assignment without ordering constraints, in which each PTA state takes the first unused equivalent candidate state. This complements the ordered structural F1.

\paragraph{Structured inefficiency analysis.}
Five categories of behavioral waste are detected with step-level localization: regression loops, blind retries, redundant steps, unnecessary exploration, and cyclic patterns. Each instance receives per-tool attribution, token-waste estimation, and a severity score. Their definitions and discriminative power are reported in Appendix~\ref{app:waste-breakdown}.

\paragraph{Combined score.}
The four signals are combined with weights optimized via grid search (step 0.05, unit-sum constraint, AUROC-maximizing on the pilot calibration set):
\begin{equation}
  f(\tau_c,\mathcal{G}) =
  0.20\cdot\Phi_{\mathrm{struct}}
  + 0.15\cdot\Phi_{\mathrm{cov}}
  + 0.30\cdot(100\cdot\Phi_{\mathrm{coh}})
  + 0.35\cdot(100\cdot\Phi_{\mathrm{temp}}).
\end{equation}
$\Phi_{\mathrm{struct}}$ and $\Phi_{\mathrm{cov}}$ are natively on a 0--100 scale; $\Phi_{\mathrm{coh}}$ and $\Phi_{\mathrm{temp}}$ are on $[0,1]$ and are rescaled by 100 so that all four terms contribute on a common scale. Behavioral signals carry 65\% of the weight. This is an empirical result of the grid search rather than a design choice. It reflects the fact that structural coverage and behavioral process quality fail on different trajectories: a failing agent that touches 70\% of ground-truth states through chaotic trial-and-error can be structurally similar to a passing agent with comparable coverage but more principled reasoning. The behavioral signals catch the difference.

\subsection{Scoring Signal Examples}
\label{app:coherence-example}

Figure~\ref{fig:scoring-examples} illustrates all four scoring signals on paired examples, contrasting a principled trajectory against a chaotic one for each dimension.

\begin{figure}[h]
  \centering
  \includegraphics[width=\linewidth,height=0.78\textheight,keepaspectratio]{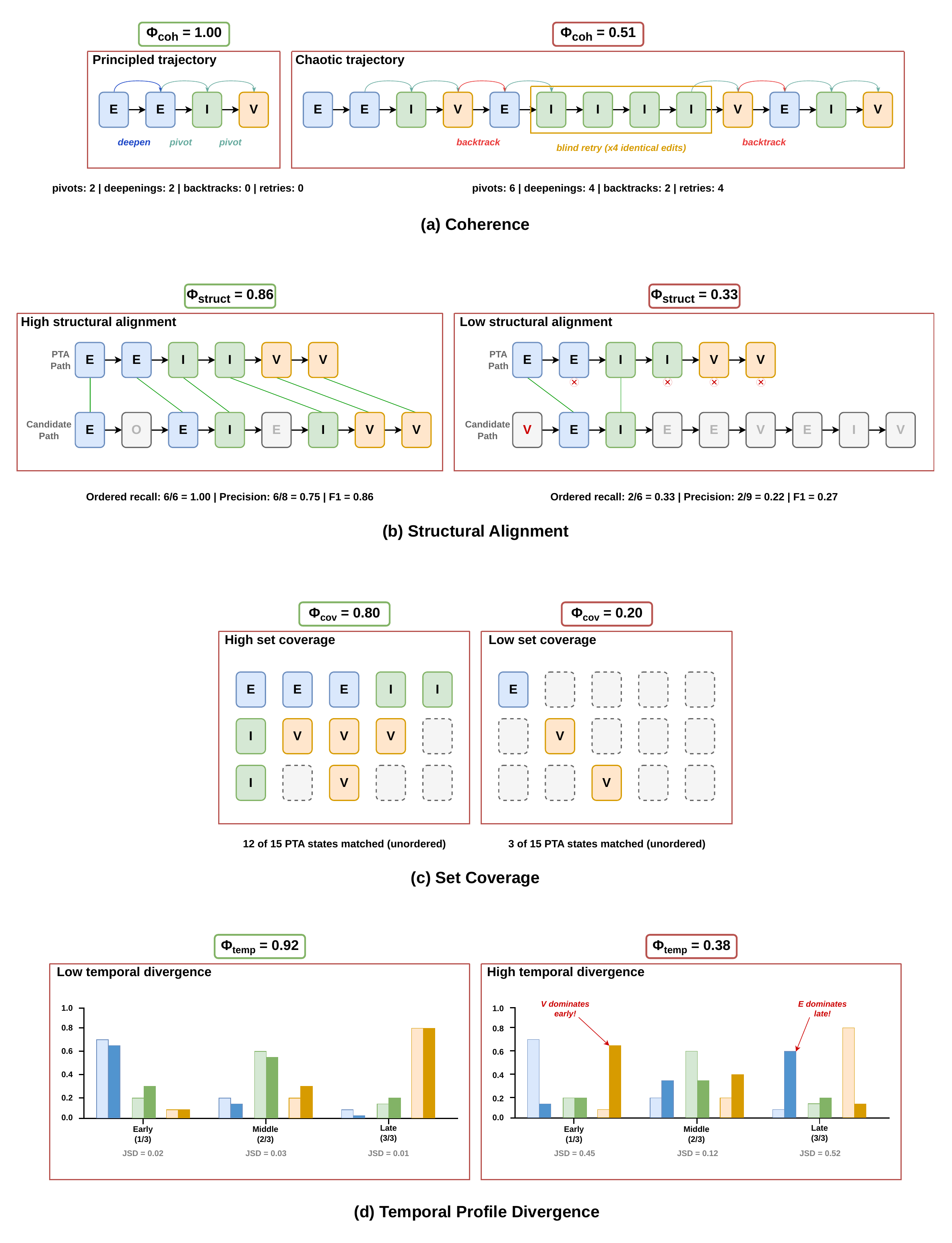}
  \caption{\textbf{Scoring signal examples.} Each panel contrasts a principled trajectory (left) against a chaotic trajectory (right) on one of the four scoring dimensions: (a)~coherence, (b)~structural alignment, (c)~set coverage, and (d)~temporal profile divergence. Together, the four signals capture complementary aspects of process quality.}
  \label{fig:scoring-examples}
\end{figure}

\paragraph{Coherence (panel a).} The principled trajectory follows a clean E$\to$E$\to$I$\to$V sequence with 2 pivots, 2 deepenings, 0 backtracks, and 0 retries, yielding $\Phi_{\mathrm{coh}} = 1.00$. The chaotic trajectory contains backtracks (V$\to$E, I$\to$E) and a cluster of 4 identical edits flagged as blind retries, producing 6 pivots, 4 deepenings, 2 backtracks, and 4 retries. The retry penalty and the reduced forward-progress ratio together bring the score to $\Phi_{\mathrm{coh}} = 0.51$.

\paragraph{Structural alignment (panel b).} The high-alignment candidate matches 6 of 6 PTA states in order (ordered recall $= 1.00$) with 6 of 8 candidate states matching (precision $= 0.75$), giving F1 $= 0.86$. The low-alignment candidate matches only 2 of 6 PTA states in order and 2 of 9 total, giving F1 $= 0.27$. The gap reflects the difference between a candidate that follows the PTA path and one that takes a substantially different route through the solution space.

\paragraph{Set coverage (panel c).} Set coverage ignores ordering and asks how many PTA states the candidate touches at all. The high-coverage candidate matches 12 of 15 PTA states ($\Phi_{\mathrm{cov}} = 0.80$). The low-coverage candidate matches only 3 of 15 ($\Phi_{\mathrm{cov}} = 0.20$). A candidate can have high coverage but low structural alignment if it visits the right states in a scrambled order, which is why both signals are needed.

\paragraph{Temporal profile divergence (panel d).} Each trajectory is divided into three equal segments (early, middle, late), and the stage distribution in each segment is compared against the PTA via Jensen-Shannon divergence. The principled trajectory has low JSD in all three segments (0.02, 0.03, 0.01), yielding $\Phi_{\mathrm{temp}} = 0.92$. The chaotic trajectory has high JSD because verification dominates the early segment (where exploration should dominate) and exploration dominates the late segment (where verification should dominate), yielding $\Phi_{\mathrm{temp}} = 0.38$. This signal catches temporal disorder that coherence alone would miss: an agent can have reasonable forward progress but do everything in the wrong order.

% ============================================================
%  APPENDIX C: ADDITIONAL EXPERIMENTAL RESULTS
% ============================================================

\section{Additional Experimental Results}
\label{app:experimental-details}

The main text reports the central experimental findings: quality stratification among passing trajectories (Section~\ref{sec:stratification}), the Lucky Pass taxonomy and model comparison (Section~\ref{sec:lucky-taxonomy}), scoring validity (Section~\ref{sec:complementarity}), and intent-label reliability (Section~\ref{sec:rq-labels}). This appendix provides the supporting tables, figures, and breakdowns that underpin those findings. We begin with the four behavioral profiles that explain why the combined score is more informative than any single signal (C.1). We then report the full pass/fail waste breakdown (C.2) and the Ideal-versus-Lucky waste comparison that identifies the blind-retry severity fingerprint (C.3). Representative failure-mode timelines (C.4) and the per-stage labeler validation (C.5) follow. Score distributions (C.6), baseline comparison (C.7), and model-comparison visualizations (C.8) complete the experimental detail.

\subsection{Behavioral Profiles}
\label{app:behavioral-profiles}

The three quality tiers (Ideal, Solid, Lucky) describe aggregate outcome categories. Within these tiers, trajectories cluster into four behavioral profiles that reveal why the combined score is more useful than any single signal. Table~\ref{tab:profiles} reports the signal ranges and prevalence for each profile.

The most instructive profile is \emph{efficient-but-atypical}, which accounts for roughly 53\% of passing trajectories. These agents follow an unconventional but internally coherent solution path: high coherence and good temporal alignment combined with low structural F1 relative to the PTA. A single-signal evaluation based on structural alignment alone would flag these as weak. The combined score correctly identifies them as Solid rather than Lucky, because the behavioral signals are high even though the structural signals are low. This is the practical payoff of multi-signal fusion: it prevents principled-but-unconventional trajectories from being confused with chaotic ones.

\begin{table}[!ht]
\centering
\caption{Behavioral profiles among passing trajectories.}
\label{tab:profiles}
\small
\begin{tabular}{@{}lccccc@{}}
\toprule
\textbf{Profile} & \textbf{Struct} & \textbf{Coh.} & \textbf{Temp.} & \textbf{Score} & \textbf{Prev.} \\
\midrule
Principled solver       & .80--.95 & .85--1.0 & .80--.95 & 75--95 & $\sim$18\% \\
Efficient, atypical     & .30--.55 & .80--1.0 & .75--.90 & 55--72 & $\sim$53\% \\
Exploratory, correct    & .55--.75 & .35--.55 & .55--.75 & 45--62 & $\sim$19\% \\
Lucky pass              & .25--.45 & .30--.50 & .35--.55 & 22--46 & $\sim$10\% \\
\bottomrule
\end{tabular}
\end{table}

\subsection{Pass/Fail Waste Breakdown}
\label{app:waste-breakdown}

Table~\ref{tab:inefficiency} provides the full pass/fail waste breakdown. Waste figures are mean steps wasted per trajectory that contains at least one instance of the category. All waste detections are ground-truth-aware: patterns already present in the merged PTA are excluded, so the numbers reflect genuinely unnecessary behavior rather than valid exploration strategies that happen to differ from one reference path.

The prevalence figures deserve careful reading. Regression loops, blind retries, and redundant steps are all slightly more common in passing trajectories than failing ones (F/P $<$ 1.0). This is a length effect rather than a detection failure: Ideal-tier passing trajectories are longer and more thorough than most failing ones, exploring more of the solution space and therefore generating more GT-excluded detections in absolute terms. The discriminating signal lies in two categories: unnecessary exploration (F/P $=$ 1.58), where failing trajectories are 58\% more likely to inspect files outside the known-good solution space, and cyclic patterns (F/P $=$ 1.32), which cost 7.8 steps per instance in failing runs versus 4.6 in passing runs.

\begin{table}[!ht]
\centering
\caption{Waste prevalence and per-instance step cost: passing (P) versus failing (F). F/P is the ratio of failing to passing prevalence; values below 1 indicate the category is more prevalent among passing trajectories.}
\label{tab:inefficiency}
\small
\begin{tabular}{@{}lccccc@{}}
\toprule
\textbf{Type} & \textbf{Prev.~P} & \textbf{Prev.~F} & \textbf{F/P} & \textbf{Waste~P} & \textbf{Waste~F} \\
\midrule
Regression loops        & 38.7\% & 39.5\% & 1.02 & 15.5 & 12.2 \\
Blind retries           & 46.0\% & 44.9\% & 0.98 & 5.6  & 7.7 \\
Redundant steps         & 50.7\% & 47.6\% & 0.94 & 3.7  & 5.3 \\
Unnecessary exploration & 6.1\%  & 9.6\%  & 1.58 & 2.0  & 1.9 \\
Cyclic patterns         & 33.6\% & 44.5\% & 1.32 & 4.6  & 7.8 \\
\bottomrule
\end{tabular}
\end{table}

\subsection{Ideal-versus-Lucky Waste}
\label{app:waste-tiers}

Table~\ref{tab:waste_tiers} reports waste by quality tier among passing trajectories. The key Lucky-pass signal is blind-retry severity: Lucky trajectories have slightly lower blind-retry prevalence, but when retries occur, they waste substantially more steps.

\begin{table}[!ht]
\centering
\caption{Waste by quality tier among passing trajectories (Ideal $n=229$, Lucky $n=122$). L/I below 1 indicates lower prevalence in Lucky.}
\label{tab:waste_tiers}
\small
\begin{tabular}{@{}lccccc@{}}
\toprule
\textbf{Type} & \textbf{Prev.~Ideal} & \textbf{Prev.~Lucky} & \textbf{L/I} & \textbf{Waste~Ideal} & \textbf{Waste~Lucky} \\
\midrule
Regression loops        & 47.6\% & 18.0\% & 0.38 & 18.6 & 7.1 \\
Blind retries           & 44.5\% & 39.3\% & 0.88 & 2.7  & \textbf{11.4} \\
Redundant steps         & 57.2\% & 24.6\% & 0.43 & 2.9  & 4.3 \\
Unnecessary exploration & 6.6\%  & 0.8\%  & 0.13 & 2.0  & 1.0 \\
Cyclic patterns         & 41.5\% & 15.6\% & 0.38 & 3.9  & 3.4 \\
\bottomrule
\end{tabular}
\end{table}

\subsection{Failure-mode Gallery}
\label{app:failure-gallery}

Figure~\ref{fig:failure-modes} gives representative timelines for the waste patterns detected by the pipeline. These examples are illustrative; the quantitative claims in the main text come from the full 1{,}815-trajectory evaluation set.

\begin{figure}[!htbp]
  \centering
  \includegraphics[width=0.88\linewidth]{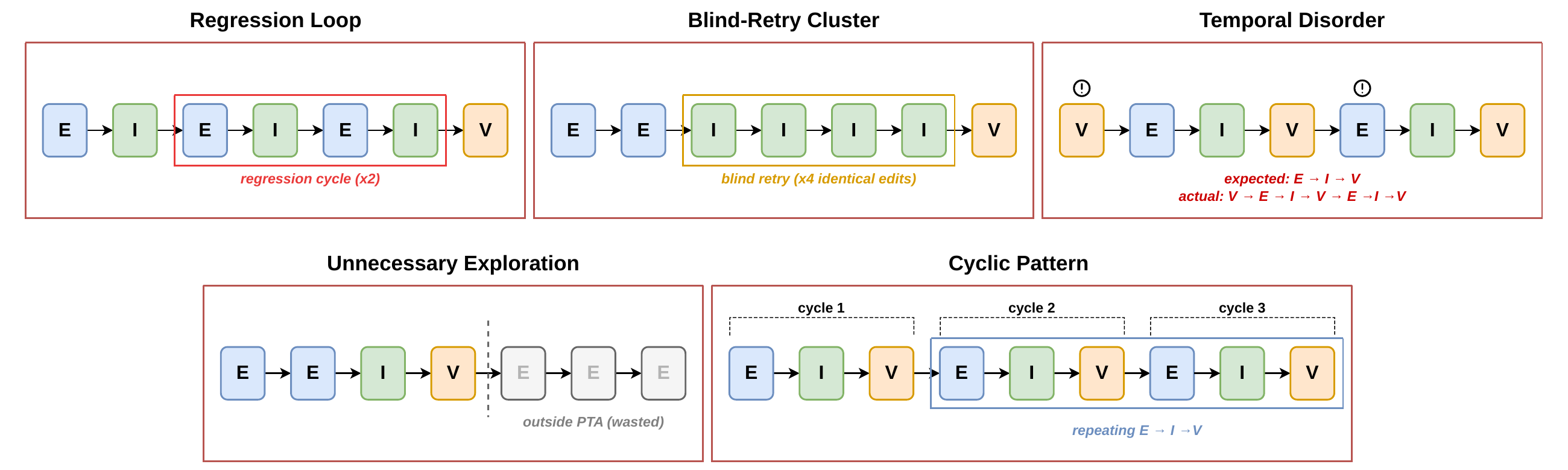}
  \caption{\textbf{Failure-mode gallery.} Six annotated stage-colored timelines: regression loop, blind-retry cluster, temporal disorder, E/V confusion before and after the context-sensitive fix, unnecessary exploration, and a cyclic pattern.}
  \label{fig:failure-modes}
\end{figure}

\subsection{Heuristic Labeler Validation}
\label{app:heuristic-validation}

The main text reports aggregate labeler reliability (Section~\ref{sec:rq-labels}): Fleiss' $\kappa = 0.933$ with 96.0\% raw agreement across seven annotators. This subsection provides the per-stage breakdown that reveals where the labeler succeeds and where it struggles.

Table~\ref{tab:kappa} gives inter-annotator agreement by category. The highest agreement is on Orchestration ($\kappa = 1.000$), which is expected since orchestration actions (thinking steps, bookkeeping) are unambiguous. The E-versus-V distinction on terminal commands, the hardest boundary, still reaches $\kappa = 0.939$. Implementation has the lowest $\kappa$ (0.713) but the highest raw agreement ($>$99\%), a statistical artifact of the small support (only 8 implementation states in the 200-state sample).

Table~\ref{tab:heuristic} evaluates the \agentlens deterministic heuristic against annotator consensus. The heuristic achieves 93.8\% accuracy and macro-F1 = 0.933. The eight E/V disagreements concentrate on post-implementation \texttt{read\_file} calls, the exact boundary case that the context-sensitive labeler targets. In these cases, the annotators labeled the read as Verification (the agent is checking its edit), while the heuristic labeled it as Exploration (the agent is reading a file). The distinction is genuinely ambiguous, and the heuristic's conservative choice (Exploration) prevents over-counting verification coverage.

\begin{table}[!ht]
\centering
\caption{Inter-annotator agreement on intent-stage labels (200 states, 7 annotators).}
\label{tab:kappa}
\small
\begin{tabular}{@{}lcc@{}}
\toprule
\textbf{Category} & \textbf{$\kappa$} & \textbf{Agreement} \\
\midrule
Overall (E/I/V/O)            & \textbf{0.933} & 96.0\% (192/200) \\
E vs.\ V (terminal commands) & 0.939          & 97.6\% \\
Implementation (I)           & 0.713          & $>$99\% \\
Orchestration (O)            & 1.000          & 100\% \\
\bottomrule
\end{tabular}
\end{table}

\begin{table}[!ht]
\centering
\caption{Heuristic classifier versus annotator consensus (192/200 states with $\geq 67\%$ agreement across 7 annotators: 2 LLMs, 5 humans).}
\label{tab:heuristic}
\small
\begin{tabular}{@{}lcccc@{}}
\toprule
\textbf{Stage} & \textbf{Precision} & \textbf{Recall} & \textbf{F1} & \textbf{Support} \\
\midrule
Exploration (E)    & 0.980 & 0.926 & 0.952 & 108 \\
Implementation (I) & 0.889 & 1.000 & 0.941 & 8 \\
Verification (V)   & 0.857 & 0.982 & 0.915 & 55 \\
Orchestration (O)  & 1.000 & 0.857 & 0.923 & 21 \\
\midrule
\textbf{Overall}   &       & 93.8\% acc & \textbf{0.933} macro-F1 & 192 \\
\bottomrule
\end{tabular}
\end{table}

\subsection{Score Distributions}
\label{app:score-distributions}

Figure~\ref{fig:score-dists-appendix} shows the pass/fail score distributions on the full 1{,}815-trajectory evaluation set. The Youden-J threshold (46.4) was computed on the pilot set and held fixed for all scaled experiments. The passing distribution concentrates in the 50--75 range with a tail extending to 95, while the failing distribution peaks around 30--40 and drops sharply above the threshold. The overlap region (roughly 35--55) contains both Solid passing trajectories and Partial-fail trajectories, which is expected: these are cases where the agent followed a reasonable strategy but either succeeded or failed on a localized implementation step. Section~\ref{sec:complementarity} reports the corresponding KS test for the combined score $f$ ($p = 0.0017$).

\begin{figure}[H]
  \centering
  \includegraphics[width=0.9\linewidth]{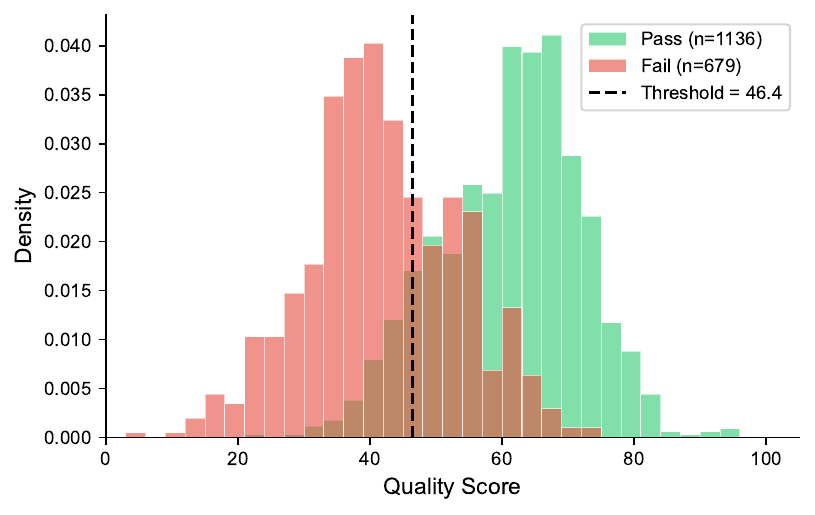}
  \caption{\textbf{Score Density by Outcome.} Overlapping density histograms of report quality scores $q$ for \textcolor{green!60!black}{Pass} ($n{=}1{,}136$) and \textcolor{red}{Fail} ($n{=}679$) instances. The vertical dashed line marks the empirically chosen threshold at 46.4. Pass instances concentrate in the upper range while fail instances skew left. Because $q$ includes the outcome term, this figure shows where the tier boundaries fall rather than testing pass/fail separation, which Section~\ref{sec:complementarity} evaluates with the combined score $f$.}

  \label{fig:score-dists-appendix}
\end{figure}

\subsection{Baseline Results}
\label{app:baseline-results}

Table~\ref{tab:baselines} compares \agentlens against three baselines. Individual matching achieves the highest AUROC, but it does not produce structured diagnostics. The PTA is designed for interpretable, actionable output rather than maximum AUROC alone.

\begin{table}[!ht]
\centering
\caption{Baseline comparison on the full 1{,}815-trajectory evaluation set.}
\label{tab:baselines}
\small
\begin{tabular}{@{}lccccc@{}}
\toprule
\textbf{Method} & \textbf{AUROC} & \textbf{Acc.} & \textbf{F1} & \textbf{KS $p$} & \textbf{API?} \\
\midrule
\agentlens (PTA)   & 0.766          & 0.720 & 0.723 & 0.0017 & No \\
Individual match.  & \textbf{0.805} & \textbf{0.744} & \textbf{0.751} & 0.0031 & No \\
\midrule
TF-IDF align.      & 0.672          & 0.651 & 0.639 & 0.182 & No \\
Dense embed.       & 0.701          & 0.667 & 0.658 & 0.094 & Yes \\
\bottomrule
\end{tabular}
\end{table}

\paragraph{Branch-localization example.}
On \texttt{astropy\_\_astropy-13236}, the merged PTA branches at state 8 where two valid solution families fork. A candidate that edits a third file outside both branches receives only a scalar similarity score from individual matching. \agentlens instead localizes the divergence to step 8, identifies the edit site as outside both PTA branches, and flags wasted exploration. This branch-aware diagnostic output is what makes the curation, model comparison, and waste reports possible.

\subsection{Model-comparison Visualizations}
\label{app:model-comparison-figs}

The main text reports model-level aggregates in Table~2 (Section~5.2). The two figures below visualize the same data to make the rank disagreements and Lucky-rate spread easier to see at a glance.

Figure~\ref{fig:pass-vs-quality} plots pass rate against mean quality score for each of the eight model configurations. If pass rate and quality were perfectly correlated, all points would lie on a monotonic curve. Instead, the scatter shows substantial disagreement: GPT-4o sits in the lower-left quadrant by pass rate but the upper-left by quality, while Opus 4.6 sits in the upper-right by pass rate but mid-right by quality. The color coding highlights the magnitude of rank divergence.

Figure~\ref{fig:lucky-rate-bar} shows the Lucky rate per model, which is the percentage of each model's passing trajectories that fall into the Lucky tier. The 46$\times$ range from Opus 4.5 (0.5\%) to GPT-4.1 (23.2\%) is the single largest behavioral gap between models and is completely invisible to pass-rate evaluation.

\begin{figure}[h]
  \centering
  \includegraphics[width=0.94\linewidth]{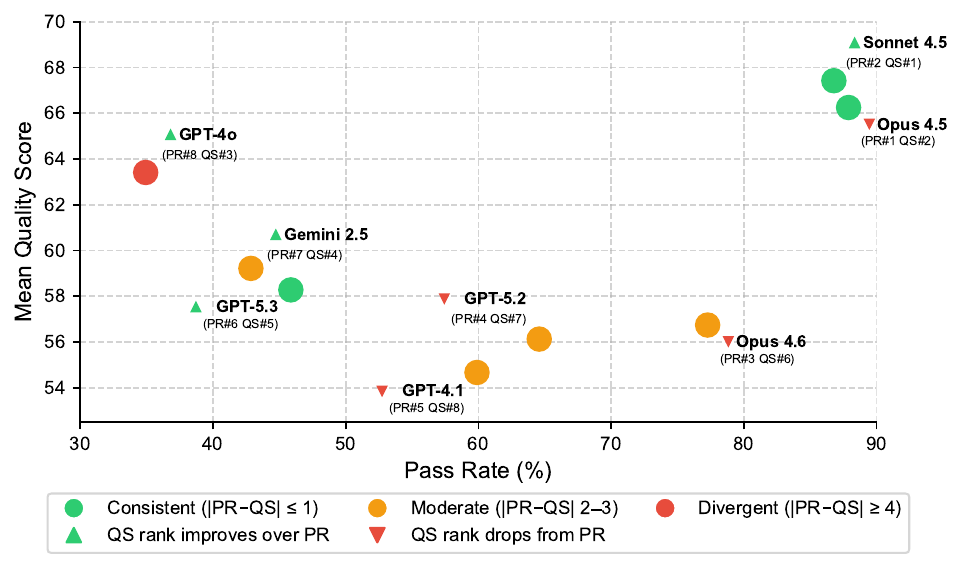}
  \caption{\textbf{Pass Rate vs.\ Mean Quality Score.} Each point represents one of eight LLM coding agents evaluated on \textsc{AgentLens-Bench}. Dot color encodes rank divergence, defined as $|\text{PR} - \text{QS}|$ where PR is the pass-rate rank and QS is the quality-score rank: \textcolor[HTML]{2ecc71}{$\bullet$}~consistent ($\leq 1$), \textcolor[HTML]{f39c12}{$\bullet$}~moderate (2--3), \textcolor[HTML]{e74c3c}{$\bullet$}~divergent ($\geq 4$). Arrows indicate whether a model's quality rank improves (\textcolor[HTML]{2ecc71}{$\blacktriangle$}) or drops (\textcolor[HTML]{e74c3c}{$\blacktriangledown$}) relative to its pass-rate rank.}
  \label{fig:pass-vs-quality}

  \vspace{0.2em}
  \includegraphics[width=0.82\linewidth]{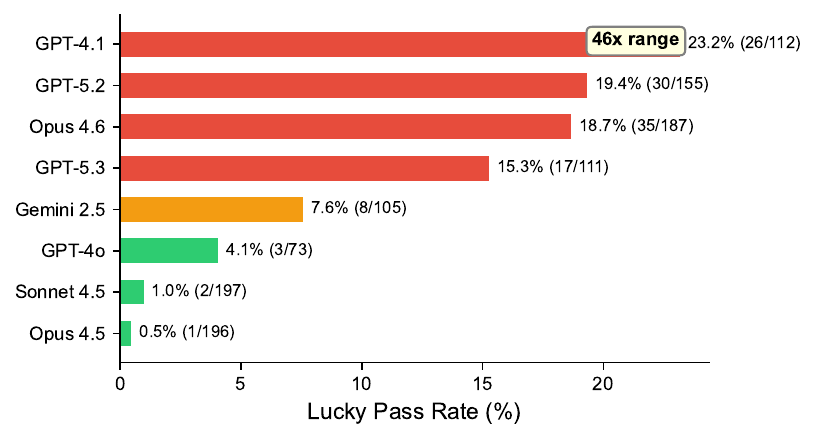}
  \caption{\textbf{Lucky rate by model.} Percentage of passing trajectories classified as Lucky tier for each model configuration. The range spans a 46$\times$ factor from Opus 4.5 (0.5\%) to GPT-4.1 (23.2\%).}
  \label{fig:lucky-rate-bar}
\end{figure}

% ============================================================
%  APPENDIX D: EXTENDED LUCKY PASS ANALYSIS
% ============================================================

\section{Extended Lucky Pass Analysis}
\label{app:lucky-analysis}

Section~\ref{sec:lucky-taxonomy} in the main text presents the Lucky Pass taxonomy, model-specific signatures, and the three-models-one-task case study. This appendix provides the full supporting analysis: the signal landscape that guided taxonomy design (D.1), the category distribution with extended descriptions (D.2), waste as a category differentiator (D.3), the complete model cross-tabulation (D.4), task-level concentration patterns (D.5), a case study showing five behavioral profiles on a single task (D.6), detailed case studies for each category (D.7), and the verification-gap analysis (D.8).

All categories are assigned automatically by the \agentlens pipeline using the structural quality signals described in Appendix~\ref{app:scoring}. No manual category labels are used. The decision tree is fully deterministic and reproducible: given the same trajectory and PTA, the same category assignment will always result.

\subsection{Signal Landscape and Taxonomy Design}
\label{app:lucky-signals}

Before defining categories, we profiled the signal distribution across all 122 Lucky-tier trajectories. Two signals are near-universal: early divergence from the ground-truth path (99.2\%) and missing verification stages (94.3\%). Because these are baseline conditions shared by almost every Lucky Pass, they cannot differentiate sub-categories. The discriminating signals are trajectory length, waste patterns, implementation completeness, and coherence.

We use a priority-ordered decision tree that assigns each trajectory to exactly one of five mutually exclusive categories. The tree checks, in order: whether the trajectory is short with zero waste and no verification (C1), whether it contains blind retries, cyclic patterns, regression loops, or thrashing (C2), whether it is long with excessive exploration (C4), whether it carries an incomplete-implementation failure reason (C3), and assigns the remainder to C5. All criteria are defined over structural quality signals computed by \agentlens.

\subsection{Category Distribution}
\label{app:lucky-category-distribution}

\begin{figure}[H]
  \centering
  \includegraphics[width=1\linewidth]{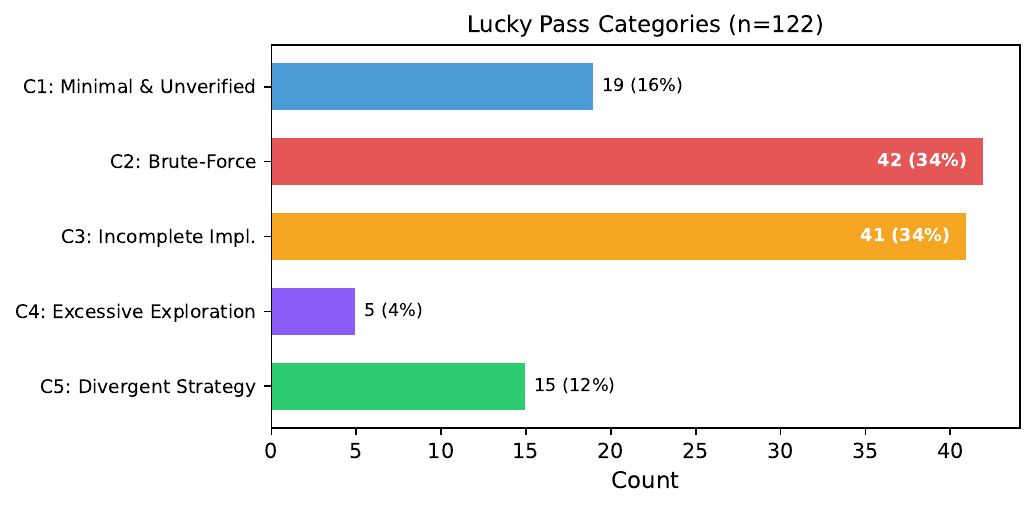}
  \caption{\textbf{Lucky Pass category distribution.} The 122 Lucky Passes decompose into five categories: C1~Minimal \& Unverified (19, 15.6\%), C2~Brute-Force Convergence (42, 34.4\%), C3~Incomplete Implementation (41, 33.6\%), C4~Excessive Exploration (5, 4.1\%), and C5~Divergent-but-Valid (15, 12.3\%). Categories C2 and C3 together account for 68\% of all Lucky Passes.}

  \label{fig:lucky-distribution-appendix}
\end{figure}

\paragraph{C1: Minimal \& Unverified ($n=19$, 15.6\%).}
The agent finds a fix in $\leq$8 steps with zero waste but skips verification entirely. These trajectories look efficient under outcome evaluation, but they lack the testing and regression checks expected from reliable SWE behavior.

\paragraph{C2: Brute-Force Convergence ($n=42$, 34.4\%).}
The agent tries multiple approaches through blind retries, cyclic patterns, or regression loops until one attempt works. This is the largest category, with mean trajectory length 35.6 states and 19.6 wasted steps per trajectory.

\paragraph{C3: Incomplete Implementation ($n=41$, 33.6\%).}
The agent implements a partial fix that addresses a surface symptom. It passes because the test suite does not cover the missing aspects of the full solution. Mean ground-truth coverage is 16.7\%, indicating that the agent overlaps with less than a fifth of the reference solution space.

\paragraph{C4: Excessive Exploration ($n=5$, 4.1\%).}
The agent explores extensively in a prolonged, unfocused manner. All five cases come from GPT-4.1, making this a model-specific termination problem.

\paragraph{C5: Divergent-but-Valid Strategy ($n=15$, 12.3\%).}
The agent takes a legitimately different approach that does not align well with the current PTA but still produces a valid fix. This is the least concerning category and reflects a known limitation of structural comparison under incomplete multi-reference coverage.

\subsection{Waste as a Category Differentiator}
\label{app:lucky-waste}

\begin{table}[!ht]
\centering
\caption{Waste metrics by Lucky Pass category. C1 and C5 both show near-zero waste but for different reasons: C1 is too short, while C5 is coherent but structurally divergent.}
\label{tab:waste-by-cat}
\small
\begin{tabular}{@{}lcccc@{}}
\toprule
\textbf{Category} & \textbf{Mean Waste} & \textbf{Severity} & \textbf{Zero-Waste \%} & \textbf{Mean Length} \\
\midrule
C1: Minimal         & 0.0  & 0.00 & 100\% & 3.2 \\
C5: Divergent-valid & 0.7  & 0.04 & 73\%  & 15.5 \\
C3: Incomplete      & 1.2  & 0.05 & 56\%  & 12.1 \\
C4: Excessive       & 3.2  & 0.09 & 20\%  & 50.4 \\
C2: Brute-Force     & 19.6 & 0.47 & 0\%   & 35.6 \\
\bottomrule
\end{tabular}
\end{table}

C1 and C5 both have near-zero waste, but for different reasons: C1 is too short to accumulate waste, while C5 often follows a coherent alternative strategy. C2 sits at the opposite extreme, with nearly half the trajectory wasted.

\subsection{Model-specific Lucky Pass Signatures}
\label{app:lucky-models}

The main text reports that Lucky Pass categories are not uniformly distributed across models (Section~\ref{sec:lucky-taxonomy}). Table~\ref{tab:lucky-model} provides the full cross-tabulation. A chi-square test on the category$\times$model contingency table yields $\chi^2(28)=102.47$, $p<0.0001$, with Cram\'er's $V=0.458$, indicating a large association. This means that knowing which model produced a Lucky Pass significantly predicts which category it falls into.

Three model-specific patterns are visible in the table. First, Opus 4.6 accounts for 89.5\% (17/19) of all C1 trajectories. It finds fixes quickly but systematically skips verification, a confidence-calibration issue where the model is skilled enough to identify correct fixes but too confident to check them. Second, GPT-4.1 dominates C2 (18/42, 43\%) and is the only model producing C4 instances. It compensates for lower initial accuracy with persistence, generating trajectories up to 100 steps long. Third, GPT-5.2-Codex and GPT-5.3-Codex cluster in C3, producing partial fixes that pass because existing test suites do not cover the full fix space. Sonnet 4.5 and Opus 4.5 produce only 2 and 1 Lucky Passes respectively, indicating that these models rarely produce low-quality passing solutions.

\begin{table}[!ht]
\centering
\caption{Lucky Pass categories by model. Bold values mark each model's dominant category.}
\label{tab:lucky-model}
\small
\begin{tabular}{@{}lcccccc@{}}
\toprule
\textbf{Model} & \textbf{C1} & \textbf{C2} & \textbf{C3} & \textbf{C4} & \textbf{C5} & \textbf{Total} \\
\midrule
opus-4.6       & \textbf{17} & 5  & 8           & 0          & 5  & 35 \\
gpt-5.2-codex  & 0           & 12 & \textbf{14} & 0          & 4  & 30 \\
gpt-4.1        & 0           & \textbf{18} & 1  & \textbf{5} & 2  & 26 \\
gpt-5.3-codex  & 2           & 1  & \textbf{12} & 0          & 2  & 17 \\
gemini-2.5-pro & 0           & 2  & \textbf{5}  & 0          & 1  & 8  \\
gpt-4o         & 0           & 2  & 1           & 0          & 0  & 3  \\
sonnet-4.5     & 0           & 2  & 0           & 0          & 0  & 2  \\
opus-4.5       & 0           & 0  & 0           & 0          & 1  & 1  \\
\bottomrule
\end{tabular}
\end{table}

Opus 4.6 accounts for 89.5\% of all C1 trajectories, indicating a tendency to find fixes quickly but skip verification. GPT-4.1 produces most C2 and all C4 trajectories, suggesting persistence without effective termination. GPT-5.2-Codex and GPT-5.3-Codex cluster in C3, producing incomplete implementations that pass visible tests.

\subsection{Task-level Concentration}
\label{app:lucky-tasks}

Of the 47 PTA-eligible tasks, 30 produce at least one Lucky Pass. The top 10 tasks account for 77 of the 122 Lucky Passes (63.1\%), and a single task, \texttt{psf\_\_requests-1724}, produces 16 Lucky Passes. A chi-square test on the category$\times$task contingency table yields $\chi^2(116)=248.45$, $p<0.0001$, with Cram\'er's $V=0.714$.

\begin{table}[!ht]
\centering
\caption{Tasks with highest Lucky Pass concentration. The top 6 tasks account for 53/122 (43.4\%) of all Lucky Passes.}
\label{tab:lucky-tasks}
\small
\resizebox{\linewidth}{!}{%
\begin{tabular}{@{}lclll@{}}
\toprule
\textbf{Task} & \textbf{Lucky} & \textbf{Categories} & \textbf{Dominant Models} & \textbf{Pattern} \\
\midrule
\texttt{psf\_\_requests-1724}       & 16 & C1:3, C2:7, C3:6 & opus-4.6, gpt-5.2-codex & All failure modes \\
\texttt{sphinx-doc\_\_sphinx-10323} & 10 & C1:1, C2:5, C5:4 & gpt-5.2-codex            & Brute-force + divergent \\
\texttt{pylint-dev\_\_pylint-6903}  & 8  & C1:5, C2:2, C3:1 & opus-4.6 (5/8)           & Overconfidence \\
\texttt{pallets\_\_flask-5014}      & 7  & C2:1, C3:6       & gpt-5.2-codex            & Incomplete impl. \\
\texttt{scikit-learn\_\_10844}      & 7  & C2:4, C3:2, C5:1 & gpt-4.1                  & Brute-force \\
\texttt{scikit-learn\_\_12585}      & 6  & C1:6             & opus-4.6 (6/6)           & Systematic overconfidence \\
\bottomrule
\end{tabular}
}
\end{table}

Lucky Pass type is a property of the model-task interaction. For example, \texttt{psf\_\_requests-1724} produces C1 from Opus 4.6, C2 from GPT-4.1, and C3 from GPT-5.3-Codex, exposing different agent shortcomings on the same task.

\subsection{Case Study: One Task, Five Behavioral Profiles}
\label{app:lucky-case-profiles}

The task \texttt{psf\_\_requests-1724} provides five passing trajectories with quality scores ranging from 22 to 88. Binary evaluation treats all rows as equivalent, while \agentlens separates an Ideal repair, a Solid repair, and three Lucky mechanisms.

\begin{table}[!ht]
\centering
\caption{Five passing trajectories for the same SWE-bench task.}
\label{tab:case-study-profiles}
\small
\begin{tabular}{@{}lrrrrrl@{}}
\toprule
\textbf{Profile} & \textbf{Model} & \textbf{Score} & \textbf{States} & \textbf{Coherence} & \textbf{Coverage} & \textbf{Category} \\
\midrule
Ideal    & GPT-4o         & 88 & 11 & 1.00 & 80.0\% & -- \\
Solid    & Opus 4.5       & 67 & 28 & 0.53 & 64.0\% & -- \\
Lucky C1 & Opus 4.6       & 33 & 4  & 0.50 & 12.0\% & Minimal and unverified \\
Lucky C3 & GPT-5.3-Codex  & 32 & 6  & 0.40 & 12.0\% & Incomplete implementation \\
Lucky C2 & GPT-4.1        & 22 & 34 & 0.03 & 16.0\% & Brute-force convergence \\
\bottomrule
\end{tabular}
\end{table}

GPT-4o produces the Ideal trajectory: it locates the relevant \texttt{method.upper()} call, implements the fix, adds a regression test, runs relevant tests, and reviews the diff. Opus 4.5 solves the task but explores more broadly than necessary. The three Lucky trajectories fail differently: Opus 4.6 stops after an unverified edit, GPT-5.3-Codex produces an incomplete implementation, and GPT-4.1 spends 34 states repeatedly reading and grepping without clear implementation progress.

\subsection{Additional Lucky Pass Cases}
\label{app:lucky-cases}

This section provides additional case studies for Lucky Pass categories. Each case study reports the task, agent, trajectory length, token cost, stage sequence, and ground-truth comparison.

\subsubsection{C2: GPT-5.2-Codex on \texttt{matplotlib\_\_matplotlib-22719}}
\label{app:c2-matplotlib}

\paragraph{Task.}
Fix matplotlib's empty category converter deprecation warning (9 GT states, 359 GT states in merged PTA).

\paragraph{Trajectory.}
18 states, 235{,}346 tokens. Steps 1--4 search the wrong directory (\texttt{src/matplotlib} instead of \texttt{lib/matplotlib}). Steps 6--11 recover by finding the correct path and reading \texttt{category.py}. Step 12 creates a reproducer script. Step 14 applies a single \texttt{replace\_string\_in\_file} edit to the \texttt{\_check\_unit} method. Steps 15--18 run the reproducer and edge-case tests.

\paragraph{Why C2.}
Two blind retries detected, trajectory thrashing (coherence 0.20), 12 wasted steps (67\% waste severity). Only 2 of 19 alignment steps match the ground truth. The wrong-directory detour and redundant file reads account for the bulk of the waste. The final edit is correct, but the path to it was not.

\paragraph{Cost context.}
235K tokens for a fix that the ground-truth solution accomplishes in 9 steps.

\subsubsection{C3: GPT-5.3-Codex on \texttt{psf\_\_requests-1724}}
\label{app:c3-requests}

\paragraph{Task.}
The same URL encoding task described in Appendix~\ref{app:lucky-case-profiles}.

\paragraph{Trajectory.}
6 states, 64{,}314 tokens. The agent lists the workspace, navigates to the testbed, reads file contents, greps for \texttt{method.upper()}, and then creates a reproducer script via \texttt{cat > repro\_unicode\_method.py}. The reproducer is labeled as ``implementation'' by the intent labeler, but it is not a source-code edit.

\paragraph{Stage sequence.}
E$\to$E$\to$E$\to$E$\to$E$\to$I (no V, no O).

\paragraph{Why C3 and not C1.}
Despite being short (6 states), this trajectory carries the \texttt{incomplete\_implementation} failure reason at high severity. The agent's implementation covers 0\% of ground-truth implementation steps because it wrote a reproducer rather than editing the actual source file. Coverage: 12.0\%, coherence: 0.40.

\paragraph{Cross-category note.}
This task (\texttt{psf\_\_requests-1724}) appears as C1 (Opus 4.6: 4 states, targeted edit, no verification), C2 (GPT-4.1: 34 states, chaotic exploration, no edit), and C3 (GPT-5.3-Codex: 6 states, incomplete implementation). The same task produces three different Lucky Pass categories depending on the model, confirming that Lucky Pass type is a property of the model-task interaction rather than the task alone.

\subsubsection{C4: GPT-4.1 on \texttt{django\_\_django-11066}}
\label{app:c4-django}

\paragraph{Task.}
Fix Django content types management to pass \texttt{using=db} to \texttt{content\_type.save()} (15 GT states, 621 GT states in merged PTA).

\paragraph{Trajectory.}
100 states, 2{,}620{,}266 tokens. The agent begins by listing locale directories (wrong area). Step 2 finds the target file via \texttt{grep}. Step 3 reads \texttt{contenttypes/management/\_\_init\_\_.py}. Step 4 attempts a \texttt{replace\_string\_in\_file} edit that fails (\texttt{old\_str} not found). Step 6 succeeds with a corrected edit (adds \texttt{using=db}). Step 43 runs the contenttypes test suite and tests pass. Steps 44--100 are 57 additional exploration steps: the agent reads the same directories repeatedly, producing identical \texttt{ls} output on steps 96--100.

\paragraph{Stage sequence.}
E$\to$E$\to$E$\to$I$\to$V$\to$I$\to$O$\to$E$\times$35$\to$V$\to$E$\times$57.

\paragraph{Cost.}
2.62M tokens for a one-line fix. This is 77$\times$ the C1 average (34K). Of the 100 states, 95 use \texttt{run\_in\_terminal}, 2 use \texttt{read\_file}, 2 use \texttt{replace\_string\_in\_file}, and 1 uses \texttt{think}. Only 4 alignment steps match.

\paragraph{Model-specific pattern.}
All five C4 instances come from GPT-4.1. The model lacks a termination heuristic: it continues exploring after the fix has been verified. The 57 post-verification exploration steps provide no value but cost over 1.5M tokens.

\subsubsection{C5: Opus 4.6 on \texttt{pylint-dev\_\_pylint-4970}}
\label{app:c5-pylint}

\paragraph{Task.}
Fix pylint's \texttt{similar} checker to handle \texttt{min\_similarity\_lines=0} (21 GT states, 276 GT states in merged PTA, 24 files).

\paragraph{Trajectory.}
8 states, 88{,}207 tokens. Six exploration steps search for the target file, read the \texttt{run()}, \texttt{process\_module}, and \texttt{\_compute\_sims} methods. Two implementation steps edit \texttt{pylint/checkers/similar.py}, modifying \texttt{process\_module} and \texttt{run()} to add early-return guards for \texttt{min\_similarity\_lines=0}.

\paragraph{Stage sequence.}
E$\to$E$\to$E$\to$E$\to$E$\to$E$\to$I$\to$I.

\paragraph{Why C5.}
The agent's approach is genuinely alternative: it modifies \texttt{process\_module} and \texttt{run()} with early-return guards, while the ground truth approaches the same problem through different code paths. Coverage is 14.3\%, but the unmatched 85.7\% is mostly verification and orchestration that the agent skipped, not implementation it missed. The agent's two matched alignment steps hit ground-truth states for implementation, confirming functional overlap.

\subsubsection{C5: GPT-5.2-Codex on \texttt{sphinx-doc\_\_sphinx-10323}}
\label{app:c5-sphinx}

\paragraph{Task.}
Fix Sphinx \texttt{literalinclude} directive's \texttt{dedent} interaction with \texttt{prepend}/\texttt{append} (53 GT states, 384 GT states in merged PTA, 60+ files).

\paragraph{Trajectory.}
14 states, 182{,}820 tokens. The agent finds the target file, reads the \texttt{append\_filter} method and filter application order, creates a reproducer, verifies the bug, implements a fix by reordering filter application in \texttt{sphinx/directives/code.py}, re-verifies, and runs a parser verification test.

\paragraph{Stage sequence.}
E$\to$E$\to$E$\to$E$\to$E$\to$E$\to$E$\to$V$\to$I$\to$V$\to$I$\to$V$\to$V$\to$O.

\paragraph{Why C5 is the least concerning category.}
Unlike the other Lucky Pass categories, this trajectory demonstrates a complete E$\to$I$\to$V$\to$O lifecycle. The agent created a reproducer, verified the bug, implemented a fix, and verified the fix works. Its low quality score (37) reflects low ground-truth coverage (11.3\%), but the coverage gap is structural (different approach) rather than qualitative (bad approach). This category argues for multi-reference ground truths and is why the PTA merges $k\geq 2$ traces.

\subsection{Verification Gap}
\label{app:lucky-verification-gap}

Lucky Passes arise from the intersection of three factors: an agent shortcoming (skipping verification, lacking planning, producing partial fixes, over-exploring, or using an alternative approach), a test-suite gap (insufficient coverage to catch the shortcoming), and task characteristics (some tasks admit multiple valid solutions). The verification gap is the most actionable of these factors because it is directly addressable through agent training or scaffolding changes.

Of the 122 Lucky Passes, 94.3\% have missing verification as a failure reason. Table~\ref{tab:lucky-verification} shows that this gap is not uniform across categories. C1 trajectories have zero verification by definition (the agent stops before running any test). C5 trajectories have the highest verification rate (87\%) because agents in this category often follow a complete E$\to$I$\to$V$\to$O lifecycle and are classified as Lucky only because their solution path diverges from the PTA. The gradient from C1 to C5 maps directly onto a gradient from most concerning to least concerning Lucky Passes.

This gradient has practical implications. Models that cluster in C1 (Opus 4.6) need verification training: the model should learn to run tests after making edits. Models that cluster in C2 (GPT-4.1) need planning and termination heuristics: the model should learn to stop exploring after a fix has been verified. Models that cluster in C3 (Codex variants) produce incomplete implementations that require better test suites to catch. The taxonomy identifies which intervention each model needs, and the verification gap is the clearest signal for prioritizing those interventions.

\begin{table}[!ht]
\centering
\caption{Verification stage coverage by Lucky Pass category.}
\label{tab:lucky-verification}
\small
\begin{tabular}{@{}lcc@{}}
\toprule
\textbf{Category} & \textbf{No verification (V=0)} & \textbf{Some verification (V$>$0)} \\
\midrule
C1: Minimal \& Unverified     & 19 (100\%) & 0 (0\%) \\
C2: Brute-Force Convergence   & 17 (40\%)  & 25 (60\%) \\
C3: Incomplete Implementation & 18 (44\%)  & 23 (56\%) \\
C4: Excessive Exploration     & 1 (20\%)   & 4 (80\%) \\
C5: Divergent-but-Valid       & 2 (13\%)   & 13 (87\%) \\
\bottomrule
\end{tabular}
\end{table}

% ============================================================
%  APPENDIX E: ABLATION DETAILS
% ============================================================

\section{Ablation Details}
\label{app:ablation-details}

Section~\ref{sec:ablation} in the main text summarizes the ablation findings in compact form. This appendix provides the full tables and extended discussion. All ablation experiments use the pilot calibration set introduced in Section~\ref{sec:setup}. The pilot contains 278 trajectories across 10 SWE-bench tasks from five agent configurations run under the OpenHands scaffold, and is entirely disjoint from the 2{,}614-trajectory scaled evaluation corpus. The pilot was used for two purposes: grid-search weight optimization (producing the weight vector $w=(0.20, 0.15, 0.30, 0.35)$ with pilot AUROC $=0.755$ and F1 $=0.791$) and the controlled ablation experiments below. All weights and thresholds were frozen before any scaled-set experiment was run.

We test three design decisions. First, whether all four scoring signals are necessary or whether a subset would suffice (E.1). Second, how many passing trajectories should be merged into each task-level PTA (E.2). Third, whether the order in which trajectories are merged affects the resulting PTA and downstream scores (E.3).

\subsection{Signal Contribution}
\label{app:signal-ablation}

The combined score fuses four signals. To test whether each signal is genuinely necessary, we ablate one at a time and measure the AUROC drop on the pilot holdout. Table~\ref{tab:ablation} reports the results.

The two behavioral signals (temporal profile and trajectory coherence) produce the largest drops when removed ($-0.037$ and $-0.031$ respectively), confirming that the behavioral dimension is not redundant with structural matching. The two structural signals produce smaller but still meaningful drops ($-0.024$ for set coverage, $-0.016$ for structural alignment). No single signal is dispensable: even the smallest drop (structural alignment) represents a statistically meaningful reduction in discrimination.

The weight vector is robust to perturbation. Shifting any single weight by $\pm 0.05$ (while re-normalizing to maintain the unit-sum constraint) reduces combined AUROC by at most 0.006. This stability means that the exact weight values are not fragile design choices, and practitioners who want to emphasize a specific dimension (e.g., weighting verification discipline more heavily for safety-critical applications) can do so without breaking the overall scoring system.

\begin{table}[!ht]
\centering
\caption{Signal ablation on the pilot calibration set ($n=278$). Behavioral signals are more impactful than structural ones in isolation; all four are necessary for maximum discrimination.}
\label{tab:ablation}
\small
\begin{tabular}{@{}lcc@{}}
\toprule
\textbf{Configuration} & \textbf{AUROC} & $\Delta$ from full \\
\midrule
Full combined (all 4 signals)                          & 0.755 & --- \\
Remove temporal profile ($\Phi_{\mathrm{temp}}$)       & 0.718 & $-0.037$ \\
Remove trajectory coherence ($\Phi_{\mathrm{coh}}$)    & 0.724 & $-0.031$ \\
Remove set coverage ($\Phi_{\mathrm{cov}}$)            & 0.731 & $-0.024$ \\
Remove structural alignment ($\Phi_{\mathrm{struct}}$) & 0.739 & $-0.016$ \\
\bottomrule
\end{tabular}
\end{table}

\subsection{Merge-count Sensitivity}
\label{app:mergecount}

The number of passing trajectories $k$ merged into each task-level PTA controls a precision-coverage trade-off. Table~\ref{tab:mergecount} reports combined AUROC as a function of $k$, evaluated on the pilot set with 3 random resamples per $k$ per eligible task.

At low $k$, the PTA is compact and precise: it encodes only a few solution strategies and penalizes any valid alternative not represented in the reference. At higher $k$, the PTA admits more valid paths and better represents the diversity of correct solutions, but the graph becomes more permissive and branches that correspond to genuine strategic choices become harder to distinguish from noise. At $k\geq 6$, a secondary effect dominates: PTA size causes some tasks to exceed scoring limits, and only easier-to-score task resamples survive. The rising AUROC at these values reflects this survivorship bias rather than genuine improvement in scoring quality.

$k=2$ achieves the highest AUROC (0.749) at full task coverage, but it encodes at most two solution strategies per task. For tasks with multiple valid approaches, this is too narrow. $k=5$ provides a better balance: it covers a substantially larger solution space, its AUROC (0.777) exceeds that of $k=2$, and the reduction in task coverage from 41 to 31 resamples reflects the PTA-size scoring limit rather than a methodological failure.

\begin{table}[!ht]
\centering
\caption{Merge-count sensitivity on the pilot set (3 resamples per $k$ per eligible task). $k=5$ balances solution-space coverage against scoring precision.}
\label{tab:mergecount}
\small
\begin{tabular}{@{}lccc@{}}
\toprule
$k$ & \textbf{AUROC} & \textbf{Acc.} & \textbf{Task-resamples scored} \\
\midrule
2 & 0.749 $\pm$ 0.232 & 81.2\% & 41 / 41 \\
3 & 0.697 $\pm$ 0.273 & 79.6\% & 40 / 41 \\
4 & 0.670 $\pm$ 0.293 & 77.2\% & 36 / 41 \\
\textbf{5} & \textbf{0.777} $\pm$ \textbf{0.220} & \textbf{83.1\%} & 31 / 41 \\
6 & 0.863 $\pm$ 0.186 & 89.7\% & 13 / 41 \\
7 & 0.822 $\pm$ 0.133 & 81.3\% & 10 / 41 \\
\bottomrule
\end{tabular}
\end{table}

We use $k=5$ for the scaled experiments because it covers a broader solution space than $k=2$ while remaining below the severe survivorship regime observed at larger $k$.

\subsection{Merge-order Robustness}
\label{app:merge-order}

Because PTA construction is incremental (trajectories are merged one at a time into the growing graph), a natural concern is whether the order in which trajectories are merged affects the resulting PTA structure and downstream scores. If it did, the scoring pipeline would be fragile: the same set of passing trajectories could produce different quality scores depending on an arbitrary processing order.

To test this, we selected one pilot task (\texttt{astropy\_\_astropy-12907}) with $k=4$, enumerated 10 random trajectory combinations, and ran all 6 permutations of each combination, yielding 60 total scoring runs. Table~\ref{tab:merge_order} reports the per-combination results. Trajectory selection accounts for 64.1\% of total variance, while merge ordering accounts for 35.9\%. Seven of ten combinations produce zero within-combination variance and are fully order-invariant. The three combinations with nonzero ordering variance (1, 5, 6) share a common property: they include at least one trajectory whose exploration prefix is ambiguous under the equivalence engine, so that small ordering differences determine whether a prefix state is merged or branched. Even in these cases, the AUROC range is bounded and the effect is substantially smaller than the effect of which trajectories are selected. The main source of PTA variation is reference-set choice, not merge ordering.

\begin{table}[!ht]
\centering
\caption{Merge-order study: per-combination AUROC across all 6 permutations ($k=4$, \texttt{astropy\_\_astropy-12907}, 60 total runs).}
\label{tab:merge_order}
\small
\begin{tabular}{@{}lccc@{}}
\toprule
\textbf{Combination} & \textbf{AUROC (mean $\pm$ std)} & \textbf{Perfect?} & \textbf{Order-invariant?} \\
\midrule
0 & $0.667 \pm 0.000$ & No  & \checkmark \\
1 & $0.778 \pm 0.172$ & No  & $\times$ \\
2 & $1.000 \pm 0.000$ & \checkmark & \checkmark \\
3 & $1.000 \pm 0.000$ & \checkmark & \checkmark \\
4 & $1.000 \pm 0.000$ & \checkmark & \checkmark \\
5 & $0.944 \pm 0.136$ & No  & $\times$ \\
6 & $0.833 \pm 0.183$ & No  & $\times$ \\
7 & $1.000 \pm 0.000$ & \checkmark & \checkmark \\
8 & $1.000 \pm 0.000$ & \checkmark & \checkmark \\
9 & $1.000 \pm 0.000$ & \checkmark & \checkmark \\
\midrule
\textbf{Overall} & $\mathbf{0.922 \pm 0.141}$ & 6/10 & 7/10 \\
\bottomrule
\end{tabular}
\end{table}

% ============================================================
%  APPENDIX F: TOKEN COST AND STATISTICAL TESTS
% ============================================================

\section{Additional Robustness Analyses}
\label{app:robustness}

Section~\ref{sec:lucky-taxonomy} in the main text compares the eight model backends, and Section~\ref{sec:pta} describes how task-level PTAs are built. This appendix provides additional analyses of both: the model comparison (Appendices~\ref{app:robust-shared}--\ref{app:robust-c5}) and PTA construction (Appendices~\ref{app:robust-lomo}--\ref{app:robust-reference-quality}). The model-comparison analyses use the published quality scores of \agentlens-Bench, the same scores as in Table~\ref{tab:model-comparison}. The PTA-construction analyses re-score the released trajectories against task-level PTAs rebuilt from the released passing trajectories. Each PTA merges $k=5$ passing trajectories of the same task, drawn in a fixed random order (seed 42); it never contains the trajectory being scored and, unless stated otherwise, never contains trajectories from the scored model. Because these PTAs differ from the merged PTAs used for the main results, absolute values differ from the main text (for example, the Lucky rate under this baseline is 8.3\%), so we report each of these analyses relative to the same baseline.

% ------------------------------------------------------------
\subsection{Model Comparison on Shared Tasks}
\label{app:robust-shared}

A model with a lower pass rate contributes fewer successful trajectories, so its quality score could reflect an easier subset of tasks. We therefore restrict the comparison to the 22 tasks solved at least once by all eight model backends (Table~\ref{tab:robust-shared}). The ordering is identical to Table~\ref{tab:model-comparison} (Kendall's rank correlation $\tau=1.0$), and GPT-4o remains third. Across all 47 tasks, task difficulty, measured as the share of failed trajectories on a task, is uncorrelated with the mean quality score of the task's passing trajectories (Pearson $r=0.00$). Because the 22 shared tasks do not span the full difficulty range, we treat this as a targeted check of task selection rather than as an alternative ranking.

\begin{table}[!ht]
\centering
\caption{\textbf{Model comparison on the 22 tasks solved by all eight backends.} The ordering matches Table~\ref{tab:model-comparison}.}
\label{tab:robust-shared}
\small
\begin{tabular}{@{}lcccc@{}}
\toprule
\textbf{Model} & \textbf{Quality (shared)} & \textbf{Rank (shared)} & \textbf{Rank (Table~\ref{tab:model-comparison})} & \textbf{Lucky\% (shared)} \\
\midrule
sonnet-4.5     & 68.0 & 1 & 1 & 0.0\%  \\
opus-4.5       & 67.7 & 2 & 2 & 0.0\%  \\
gpt-4o         & 63.8 & 3 & 3 & 4.7\%  \\
gemini-2.5-pro & 59.9 & 4 & 4 & 6.3\%  \\
gpt-5.3-codex  & 59.8 & 5 & 5 & 11.3\% \\
opus-4.6       & 58.4 & 6 & 6 & 17.5\% \\
gpt-5.2-codex  & 57.4 & 7 & 7 & 17.8\% \\
gpt-4.1        & 55.5 & 8 & 8 & 21.4\% \\
\bottomrule
\end{tabular}
\end{table}

% ------------------------------------------------------------
\subsection{Overall Quality per Attempt}
\label{app:robust-per-attempt}

The quality score in Table~\ref{tab:model-comparison} is computed over successful trajectories and describes how cleanly a model works when it succeeds (a conditional quality score). To describe process quality across all attempts, we also report the mean quality score over every trajectory of a model,
\begin{equation}
  \mathbb{E}[Q] = p_{\mathrm{pass}}\,\mathbb{E}[Q \mid \mathrm{pass}] + (1-p_{\mathrm{pass}})\,\mathbb{E}[Q \mid \mathrm{fail}],
\end{equation}
where $p_{\mathrm{pass}}$ is the model's pass rate. All eight backends are evaluated on the same 47 tasks. Because $q$ includes the outcome term $O$, this average adds $10\,p_{\mathrm{pass}}$ points to each model, so it combines process quality with success frequency (Table~\ref{tab:robust-per-attempt}). Under this measure, GPT-4o moves from third to fourth, below Opus 4.6, and the ranking agrees only moderately with the pass-rate ranking (Kendall's $\tau=0.43$). Without the outcome term, GPT-4o remains third (49.3, compared with 45.8 for Opus 4.6), because its failing trajectories also receive higher quality scores (47.1, compared with 37.7 for GPT-5.3-Codex and 37.4 for GPT-5.2-Codex).

\begin{table}[!ht]
\centering
\caption{\textbf{Overall Quality per Attempt.} Mean quality score over all trajectories of each model, combining success frequency with the quality of passing and failing trajectories.}
\label{tab:robust-per-attempt}
\small
\begin{tabular}{@{}lccccc@{}}
\toprule
\textbf{Model} & \textbf{Pass\%} & \textbf{Quality (pass)} & \textbf{Quality (fail)} & \textbf{Quality per attempt} & \textbf{Rank} \\
\midrule
sonnet-4.5     & 86.8\% & 67.4 & 57.7 & \textbf{66.1} & 1 \\
opus-4.5       & 87.9\% & 66.2 & 55.5 & 64.9 & 2 \\
opus-4.6       & 77.3\% & 56.7 & 42.4 & 53.5 & 3 \\
gpt-4o         & 34.9\% & 63.4 & 47.1 & 52.8 & 4 \\
gpt-5.2-codex  & 64.6\% & 56.1 & 37.4 & 49.5 & 5 \\
gemini-2.5-pro & 42.9\% & 59.2 & 39.3 & 47.8 & 6 \\
gpt-5.3-codex  & 45.9\% & 58.3 & 37.7 & 47.1 & 7 \\
gpt-4.1        & 59.9\% & 54.7 & 34.6 & 46.6 & 8 \\
\bottomrule
\end{tabular}
\end{table}

% ------------------------------------------------------------
\subsection{Lucky Rate Without Divergent-but-Valid Passes}
\label{app:robust-c5}

Category C5 (Divergent-but-Valid, Table~\ref{tab:lucky-taxonomy-main}) contains Lucky Passes that are coherent and correct but structurally different from the task-level PTA. Their low scores partly reflect incomplete PTA coverage rather than weak agent behavior. Table~\ref{tab:robust-c5} reports the Lucky rate with and without C5, using the category counts in Table~\ref{tab:lucky-model}. Excluding C5 lowers the overall Lucky rate from 10.7\% to 9.4\%, and the order of models by Lucky rate is unchanged.

\begin{table}[!ht]
\centering
\caption{\textbf{Lucky rate including and excluding C5 (Divergent-but-Valid).} Percentages are over each model's passing trajectories; C5 counts are from Table~\ref{tab:lucky-model}.}
\label{tab:robust-c5}
\small
\begin{tabular}{@{}lcccc@{}}
\toprule
\textbf{Model} & \textbf{Passing} & \textbf{C5} & \textbf{Lucky\% (all)} & \textbf{Lucky\% (excl.\ C5)} \\
\midrule
sonnet-4.5     & 197 & 0 & 1.0\%  & 1.0\%  \\
opus-4.5       & 196 & 1 & 0.5\%  & 0.0\%  \\
gpt-4o         & 73  & 0 & 4.1\%  & 4.1\%  \\
gemini-2.5-pro & 105 & 1 & 7.6\%  & 6.7\%  \\
gpt-5.3-codex  & 111 & 2 & 15.3\% & 13.5\% \\
opus-4.6       & 187 & 5 & 18.7\% & 16.0\% \\
gpt-5.2-codex  & 155 & 4 & 19.4\% & 16.8\% \\
gpt-4.1        & 112 & 2 & 23.2\% & 21.4\% \\
\midrule
All            & 1{,}136 & 15 & 10.7\% & 9.4\% \\
\bottomrule
\end{tabular}
\end{table}

% ------------------------------------------------------------
\subsection{Same-model Trajectories in the PTA}
\label{app:robust-lomo}

Excluding the scored trajectory from its PTA does not prevent a model from benefiting when other trajectories it generated are part of that PTA. We compare two ways of building the PTA: from any model's passing trajectories (excluding only the trajectory being scored), and leave-one-model-out (LOMO), which excludes every trajectory from the scored model (Table~\ref{tab:robust-lomo}). Under LOMO, the quality score of seven of the eight backends decreases (mean absolute change 1.99 points), consistent with a small advantage from same-model trajectories in the PTA. The ranking is largely preserved ($\tau=0.86$): GPT-4o is third under LOMO, and Opus 4.6 does not improve when its own trajectories are excluded. Same-model trajectories therefore do not drive the model comparison.

\begin{table}[!ht]
\centering
\caption{\textbf{Same-model trajectories in the PTA.} Quality score when the PTA may include other trajectories from the scored model, and under leave-one-model-out (LOMO) PTAs.}
\label{tab:robust-lomo}
\small
\begin{tabular}{@{}lccccc@{}}
\toprule
\textbf{Model} & \textbf{Quality (any model)} & \textbf{Quality (LOMO)} & \textbf{Change} & \textbf{Rank (any)} & \textbf{Rank (LOMO)} \\
\midrule
sonnet-4.5     & 67.0 & 65.5 & $-1.5$ & 3 & 1 \\
opus-4.5       & 67.5 & 63.7 & $-3.8$ & 1 & 2 \\
gpt-4o         & 67.4 & 63.7 & $-3.7$ & 2 & 3 \\
gemini-2.5-pro & 60.7 & 59.4 & $-1.3$ & 4 & 4 \\
gpt-5.3-codex  & 60.4 & 58.1 & $-2.3$ & 5 & 5 \\
opus-4.6       & 59.0 & 56.3 & $-2.8$ & 6 & 6 \\
gpt-5.2-codex  & 56.1 & 55.6 & $-0.5$ & 7 & 7 \\
gpt-4.1        & 54.5 & 54.5 & $0.0$  & 8 & 8 \\
\bottomrule
\end{tabular}
\end{table}

% ------------------------------------------------------------
\subsection{Sensitivity to State-equivalence Thresholds}
\label{app:robust-thresholds}

The state-equivalence engine (Appendix~\ref{app:equivalence}) uses two tunable thresholds: the line-range overlap required to merge reads or edits of the same file (default 0.3) and the word-level Jaccard similarity required to merge search commands (default 0.5). We vary the overlap threshold from 0.1 to 0.5 and the Jaccard threshold from 0.3 to 0.7, one at a time (Table~\ref{tab:robust-thresholds}). Across all settings, the Lucky rate is unchanged, pass/fail AUROC of the combined score (Eq.~1) changes by less than 0.001, and the model ranking is identical ($\tau=1.0$). Even at the extreme settings, the PTA size is unchanged for 27 of the 47 tasks: raising both thresholds to 1.0 changes no PTA, and lowering both to 0.0 changes the remaining PTAs by a median of one state. The reported conclusions therefore do not depend on these threshold choices; which trajectories are merged into the PTA (Appendices~\ref{app:robust-lomo} and~\ref{app:robust-reference-quality}) is the more important source of PTA variation.

\begin{table}[!ht]
\centering
\caption{\textbf{Sensitivity to state-equivalence thresholds.} $\Delta$AUROC and Kendall's $\tau$ are relative to the default setting.}
\label{tab:robust-thresholds}
\small
\begin{tabular}{@{}lccc@{}}
\toprule
\textbf{Setting} & \textbf{Lucky\%} & \textbf{$\Delta$AUROC} & \textbf{Kendall's $\tau$} \\
\midrule
Default (overlap 0.3, Jaccard 0.5) & 8.3\% & ---          & ---  \\
Overlap 0.1                        & 8.3\% & $<0.001$     & 1.00 \\
Overlap 0.5                        & 8.3\% & $<0.001$     & 1.00 \\
Jaccard 0.3                        & 8.3\% & $<0.001$     & 1.00 \\
Jaccard 0.7                        & 8.3\% & $<0.001$     & 1.00 \\
\bottomrule
\end{tabular}
\end{table}

% ------------------------------------------------------------
\subsection{Lucky Passes in the PTA}
\label{app:robust-reference-quality}

Any passing trajectory can be merged into a task-level PTA (Section~\ref{sec:pta}), so a Lucky Pass can become part of the reference. At least one Lucky Pass is among the passing trajectories of 30 of the 47 tasks (63.8\%). We rebuild the PTAs without Lucky-tier trajectories (published quality score below 47); every task still has at least two eligible passing trajectories. Without Lucky Passes in the PTA, the Lucky rate rises from 8.3\% to 9.2\% ($+0.9$ percentage points), the model ranking is unchanged ($\tau=1.0$), and pass/fail AUROC changes by $-0.006$. Lucky Passes make a PTA slightly more permissive, so the reported Lucky rates are conservative with respect to the quality of the trajectories merged into the PTA.

\section{Token Cost and Statistical Tests}
\label{app:token-cost}

The Lucky Pass taxonomy reveals not only qualitative differences in agent behavior but also quantitative differences in computational cost. Binary evaluation treats all correct patches as equivalent, but the token expenditure behind those patches varies by a factor of 40 across Lucky Pass categories. Table~\ref{tab:token-cost} reports the breakdown.

The cost gradient follows a clear pattern. C1 (Minimal \& Unverified) trajectories are the cheapest because they are the shortest: mean 34K tokens per trajectory. C3 (Incomplete Implementation) and C5 (Divergent-but-Valid) are moderately expensive. C2 (Brute-Force Convergence) is 25$\times$ more expensive than C1 because the agent spends many steps retrying failed approaches. C4 (Excessive Exploration) is the most expensive at 40$\times$ C1, with one trajectory consuming 2.62M tokens for a one-line fix.

\begin{table}[!ht]
\centering
\caption{Token cost by Lucky Pass category. C4 costs 40$\times$ more than C1 on average for the same binary evaluation outcome.}
\label{tab:token-cost}
\small
\begin{tabular}{@{}lcccc@{}}
\toprule
\textbf{Category} & $n$ & \textbf{Mean Tokens} & \textbf{Range} & \textbf{vs.\ C1} \\
\midrule
C1: Minimal         & 19 & 34{,}032     & 22K--75K    & 1.0$\times$ \\
C3: Incomplete      & 41 & 162{,}024    & 55K--520K   & 4.8$\times$ \\
C5: Divergent-valid & 15 & 232{,}854    & 59K--882K   & 6.8$\times$ \\
C2: Brute-Force     & 42 & 856{,}512    & 120K--3.17M & 25.2$\times$ \\
C4: Excessive       & 5  & 1{,}377{,}226 & 569K--2.62M & 40.5$\times$ \\
\bottomrule
\end{tabular}
\end{table}

Aggregate token expenditure across all 122 Lucky Passes is approximately 58.3M tokens. If all were solved at C1 efficiency (34K each), the total would be 4.1M tokens. The excess 54M tokens (93\% of the aggregate) is attributable to C2 and C4 patterns. At current API pricing, the 122 Lucky Passes cost roughly 14$\times$ what they would cost if every model solved tasks at C1 efficiency. A model that produces C2 (brute-force) Lucky Passes is 25$\times$ more expensive per trajectory than one that produces C1 (minimal) Lucky Passes, yet both receive identical binary scores. Token efficiency is invisible to pass/fail evaluation but directly affects deployment economics.

\subsection{Statistical Tests for Lucky Pass Taxonomy}
\label{app:lucky-stats}

\begin{table}[!ht]
\centering
\caption{Association tests for the Lucky Pass taxonomy.}
\label{tab:lucky-stats}
\small
\begin{tabular}{@{}lcccc@{}}
\toprule
\textbf{Test} & $\chi^2$ & \textbf{dof} & $p$ & \textbf{Cram\'er's $V$} \\
\midrule
Category $\times$ Model & 102.47 & 28  & $<0.0001$ & 0.458 (large) \\
Category $\times$ Task  & 248.45 & 116 & $<0.0001$ & 0.714 (very large) \\
\bottomrule
\end{tabular}
\end{table}

Both associations are highly significant with large to very large effect sizes, confirming that (1) model choice strongly predicts the type of Lucky Pass produced, not just whether one occurs, and (2) task characteristics strongly predict which Lucky Pass category dominates. The per-category verification coverage further supports the causal structure described in Section~\ref{app:lucky-verification-gap}: C1 trajectories have zero verification coverage by definition, C2 and C3 have 40--60\% verification rates, and C5 has 87\% verification.

\end{document}